\documentclass[apj]{emulateapj}
\usepackage{apjfonts}
\usepackage{multirow}
\usepackage{color}
\usepackage{natbib}
\usepackage{amssymb, amsmath}
\usepackage{gensymb}
\newcommand{\beq}{\begin{equation}}
\newcommand{\eeq}{\end{equation}}
\newcommand{\beqa}{\begin{eqnarray}}
\newcommand{\eeqa}{\end{eqnarray}}

\newcommand{\rattk}{L$_{24\mu m}$/L$_K$}

\shorttitle{Virgo-COMA}
\shortauthors{Riguccini et al.}

\begin{document}
\title{Mid-IR Enhanced Galaxies in the Coma \& Virgo Cluster: lenticulars with a high star formation rate}
\author{
L. Riguccini\altaffilmark{1,2,3},
P. Temi\altaffilmark{3},
A. Amblard,\altaffilmark{2,3},
M. Fanelli\altaffilmark{2,3},
F. Brighenti\altaffilmark{4}
}

\email{riguccini@baeri.org}

\altaffiltext{1}{Observat\'orio do Valongo, Universidade Federal do Rio de Janeiro, Ladeira do Pedro Ant\^onio 43, Sa\'ude,
Rio de Janeiro, RJ 20080-090, Brazil (riguccini@astro.ufrj.br)}
\altaffiltext{2}{NASA Ames Research Center, Moffett Field, CA, USA}
\altaffiltext{3}{BAER Institute, Sonoma, CA, USA}
\altaffiltext{4}{Astronomy Department, University of Bologna, Via Ranzani 1, 40127 Bologna, Italy}

\begin{abstract}

We explore the properties of early-type galaxies (ETGs), including ellipticals (E) and lenticulars (S0), in rich environments such as clusters of galaxies (Virgo and Coma).
The L$_{24}$/L$_K$ distribution of ETGs in both Virgo and Coma clusters shows that some S0s have a much larger L$_{24}$/L$_K$ ratio (0.5 to $\sim$2 dex) than the bulk of the ETG population. This could be interpreted as an enhanced star formation rate in these  lenticulars.
We compare the optical colors of galaxies in these two clusters and investigate the nature of these sources with a
large L$_{24}$/L$_K$ ratio by looking at their spatial distribution within the cluster, by analyzing their optical spectra and
by looking at their optical colors compared to late-types.
We obtain 10 Coma and 3 Virgo early-type sources with larger L$_{24}$/L$_K$ ratios than the bulk of their population. We call these sources Mid-Infrared Enhanced Galaxies (MIEGs).
In Coma, they are mostly located in the South-West part of the cluster where a substructure is falling onto the main cluster. MIEGs present lower g-r color than the rest of the ETG sample, because of a blue continuum. We interpret the excess L$_{24}$/L$_K$ ratio as evidence for an enhanced star-formation induced as a consequence of their infall into the main cluster. 


\end{abstract}

\date{Accepted 2015 July 7}

\section{Introduction}

The stellar content of galaxies is known to vary drastically with the environment. 
Clusters of galaxies are places of prime interest to study galaxy evolution as they offer a diverse set of physical conditions: virialized regions, merging substructures, etc. To a further extend, clusters of galaxies, being at the nodes of the filamentary structures of the Universe, are the ideal locations to understand the process of matter accretion from the surrounding large-scale structures.

Early-type galaxies (ETGs), elliptical and lenticular galaxies,  used to be seen as a ``red and dead'' population, bereft of star-formation \citep[e.g.,][]{Kormendy:89}. Recent studies have indicated that some ETGs show signs of a current star-formation activity \citep[e.g.,][]{dezeeuw:02,Young:09,Shapiro:10,Amblard:14}. 
Different studies have highlighted differences in the ETG stellar population in dense cluster and in lower density environment \citep[e.g.,][]{Bower:90,Renzini:06}, based on certain spectral line ratios in these galaxies. \citet{Caldwell:93} found similar evidence, arguing that a basic visual inspection of ETG spectra was already capable of distinguishing differences in the stellar populations of ETGs lying in distinct environments (looking at Balmer absorption and emission lines).
In this paper, we choose to study ETGs in a dense environment, in two nearby galaxy clusters, that are at a different stage of their evolution,  the Virgo and Coma clusters.
From statistical analyses of the positions and velocities of individual galaxies or through X-ray maps, significant substructures have been identified in these clusters underlying the fact that they are not virialized yet \citep[e.g.][]{Fitchett:87,Geller:82,Oegerle:01,White:93,Briel:01,Arnaud:01,Bohringer:94,Binggeli:93}.\\ 
At a distance of 16.5\,Mpc \citep{Mei:07}, Virgo is the closest galaxy cluster  and is known to have a complex morphology with several components \citep[e.g.][]{deVaucouleurs:61,Pierce:88,Yasuda:97,Federspiel:98,Binggeli:87,Ftaclas:84,Yoon:12}. 
In a simplified way, Virgo is composed of 3 main substructures (cluster A with M87, cluster B with M49 and a sub-cluster of cluster A with M86), 3 clouds further away from the main cluster (M,W and W') and the southern extension. 
These substructures are interconnected and fall into one another. They are located in the far outskirts of the Virgo Cluster, outside of the virial radius. They represent a great environment to study cosmic filaments linked to the large scale structure feeding into the cluster. 

The Coma Cluster, located at a distance of 99\,Mpc, has a complex morphology given that it is undergoing a merger, with a bright subcluster located around the south-west (SW) part of the cluster center \citep{Adami:05}. The Coma cluster has two central galaxies with similar brightness, instead of a single dominant  galaxy as often is the case in clusters, a sign that  Coma has undergone at least a relatively recent major merger. Coma has more galaxies, is more concentrated and is more virialized than Virgo \citep[e.g.,][]{Weinzirl:14,Pimbblet:14,Smith:12}.
\citet{Simionescu:13} showed that the ongoing merger in the SW direction affects the surface radio profile of the cluster in this direction and that the intracluster medium (ICM) temperature along the eastern arm is systematically lower than in the north-west (NW) and west (W) directions from the center of the cluster out to 50 arcmin (1.4 Mpc).
Using data from the X-ray satellite {\em Suzaku} launched in 2005, they also detected a boost in the surface brightness distribution in the 0.7-7\,keV band towards the SW direction, location of an ongoing merger. They detect a X-ray emission out to a larger radii in the SW direction (as far as $\sim$2.5\,Mpc) than towards the other directions.

\citet{Simionescu:13} and \citet{Urban:11} have produced similar studies (primarily thermodynamical) in Coma and Virgo respectively, which give an opportunity to compare the evolution of the two clusters. They found a similar trend for the ICM temperature, which decreases with the radius, while the ICM metallicity shows a different behavior in the outskirts of the clusters. \citet{Urban:11} observed a peak in the metallicity radial distribution at the outskirts of the Virgo cluster ($\simeq$ 0.4 Mpc), whereas the metallicity of the Coma cluster mainly decreases with the radius \citep{Simionescu:13}.
The motivation of this present paper is to draw a comparison of the different mid-infrared (mid-IR) properties of the ETG population of the Virgo and Coma clusters. 
The paper is organized as follows. In section 2 we describe the data used in this work for both the Virgo and the Coma clusters. In section 3 we highlight the bimodality in the L$_{24}$/L$_{K}$ distributions between Virgo and Coma. Virgo and Coma ETGs, that shows signs of enhanced star-formation, are selected and described in section 4 and 5, analyses in section 6.  Section 7 briefly summarized our results. Throughout this paper, we adopt the standard $\Lambda$CDM cosmology: $\Omega_m$\,=\,0.3, $\Omega_{\Lambda}$\,=\,0.7 and H$_0$\,=\,70\,km\,s$^{-1}$\,Mpc$^{-1}$.

\section{Data}
\label{sec:data}

We base our work on early-type galaxies in the Virgo and Coma clusters, simultaneously detected at 24$\mu$m with the MIPS instrument onboard {\it Spitzer}, and in the $Ks$-band with the 2 Micron All-Sky Survey (2MASS).  
Morphological classifications are based on the T parameter,  obtained from the Hyperleda database\footnote{http://leda.univ-lyon1.fr/} \citep{Paturel:03}. We distinguish ellipticals with $ - 5 \le T < - 3$ from lenticulars for which $ -3 \le T < 0$. 

\subsection{Virgo sample}

The Virgo sample was assembled by cross identification between the Virgo Cluster Catalog \citep[VCC,][]{Binggeli:85} and the \citet{Temi:09b} sample, resulting in the selection of 58 galaxies with both Ks-band and 24$\mu$m emission. Eight additional galaxies from the catalog of \citet{Leipski:12} were added to produce our Virgo Cluster sample. The total sample contains 66 sources evenly split between ellipticals (N = 34) and lenticulars  (N = 32).  All Virgo galaxies have high quality Ks-band photometry from 2MASS.

Our dataset comprises photometry collected from a number of {\em Spitzer} observing programs and is not derived from a single large flux-limited sample. To assess how representative this sample is when compared to the entire Virgo Cluster ETG population, we matched our sample against the Virgo Cluster HST-ACS survey and also against the early-type galaxy component of the Virgo Cluster Catalog.  The ACS Virgo Cluster Survey \citet{Cote:04} obtained high-resolution imaging in two bandpasses (F475W and F850LP) for 100 early-type cluster members. 

Our sample contains 24 of 25 ETGs brighter than $B_{T} \le 12.0$ (96\%), 41 of 48 brighter than 13.0 (85\%) and 54 of 78 brighter than 14.0 (69\%).   Based on the classifications provided in the VCC, the Virgo Cluster contains approximately equal numbers of bright E and S0 systems, similar to the the distribution in our mid-IR selected sample. We conclude that our sample should be representative of the bright ETGs in the Virgo Cluster.  
In constructing the samples to be compared, we wish to include only those sources in which the mid-IR luminosity is derived exclusively from dust irradiated by starlight. Mid-IR fluxes can be boosted by radiatively warmed dust in accretion tori around active supermassive black holes (SBMHs) in the galaxy core.  This AGN-driven excess mid-IR luminosity can lead to an overestimate of the star formation rate of the host galaxy. To exclude galaxies contaminated by AGN-heated dust, we utilize a BPT diagram diagnostic \citep{BPT:81} to identify likely AGNs. From  the SDSS spectroscopic catalog, we found 30 counterparts for our 66 Virgo sources: 9 of the 30 sources are star-forming, one showed emission line ratios consistent with a LINER designation (NGC 4168) and 20 are unclassifiable. Additionally, we examined the literature for known Virgo AGNs, adding NGC 4486 (M87), NGC 4374 (M84) and NGC 4261 to the excluded list. 

The final Virgo sample is composed of 30 elliptical galaxies and 32 lenticulars. Four systems were excluded as AGNs, all ellipticals, whose 24$\mu$m emission is expected to be dominated by heating from the central source.  In this context we note the results of \citet{Leipski:12} who examined low-level nuclear activity in Virgo Cluster early-type galaxies by measuring the 2D 24$\mu$m light distribution in these systems. They found little evidence for strongly enhanced 24$\mu$m emission from a central AGN and concluded that SMBHs are generally quiescent, a consequence of low gas accretion rates.

\subsection{Coma sample}

For the Coma Cluster, we utilize the work of \citet{Mahajan:10}  to build our ETG sample. \citet{Mahajan:10} used a combination of MIPS 24$\mu$m observations, SDSS photometry and spectra to investigate the star formation history of galaxies in the Coma supercluster. All their galaxies from the SDSS data in the Coma supercluster region are brighter than $ r \sim 17.77$, the completeness limit of the SDSS spectroscopic galaxy catalog. Their 24$\mu$m fluxes are obtained from archival data covering $2 \times$2 deg$^2$ for Coma Cluster. \citet{Bai:06} derived a 24$\mu$m luminosity function of Coma using the same data and estimated the 80\% completeness limit of the sample to be 0.33\,mJy. They construct  a catalog of 197 sources with SDSS and 24$\mu$m counterparts within $\sim$2h$^{-1}$\,Mpc of the centre of Coma that we use to build our sample. Among the 197 sources, 134 galaxies are classified as ETGs, 38 have a morphological index of a late-type galaxy ($ T > 0$) and 25 sources do not have a morphological parameter. 124 sources among  the 134 ETGs have been observed by 2MASS in the $Ks$-band. Our final sample of 124 sources is composed of 49 ellipticals and 75 lenticulars. 

Adopting the same approach as for the Virgo sample, we choose to exclude systems with a strong AGN signal from our Coma dataset. Among the 124 ETGs of our Coma sample, 14 are flagged as AGN by \citet{Mahajan:10} using BPT diagram diagnostics or the \citet{Miller:03} criteria (1 elliptical and 13 lenticulars).  We note here the opposite trend than that observed in Virgo, where all of the flagged AGNs are classed as ellipticals. The final Coma sample is composed of 48 elliptical and  62 lenticular galaxies, when excluding 1 elliptical and 13 lenticulars classified as AGNs.

\subsection{Completeness and selection effect of the samples}

One of the main focus of this present work is a comparison between the ETGs in the Coma and Virgo clusters. Since the clusters differ in distance by a factor of 6, the observations, that we use, do not reach the same depth. To avoid  being affected by biases, we discuss the limiting fluxes of the two samples hereafter. The Coma catalog from \citet{Mahajan:10} is 80\% complete down to $F_{24\mu m}=$0.33\,mJy. Once we apply this completeness limit to our Coma selection (\citet{Mahajan:10} catalog contains galaxies below  $F_{24\mu m}=$0.33\,mJy), the sample is reduced to 95 non-AGN sources, distributed between 42 elliptical and 53 lenticular galaxies.
The Coma 80\% completeness flux at 24 $\mu$m corresponds to a 24$\mu$m flux of 12 mJy at 16.5 Mpc (Virgo) and to a 24$\mu$m luminosity of 10$^{40.7}$ erg.s$^{-1}$. The Coma completeness in the 2MASS K$_{s}$ and SDSS r band corresponds to depths at Virgo distance of K$_{s} <$ 10.53 [Vega]  (i.e. F$_{K_s} >$ 40.8 mJy) and r $<$ 12.9 [AB] (i.e. F$_r >$ 25.3 mJy). Applying this cut (the 24 $\mu$m 
completeness being the most stringent), the final Virgo sample is composed of 25 non-AGN sources: 10 elliptical galaxies and 15 lenticulars. For the remainder of this paper we consider only sources down to the same depth. 

\section{The L$_{24\mu m}$/L$_K$ distribution of ETGs in Virgo and Coma}
\label{sec:L24LK}

\begin{figure*}
 \centering
 \resizebox{1.\hsize}{!}{\includegraphics{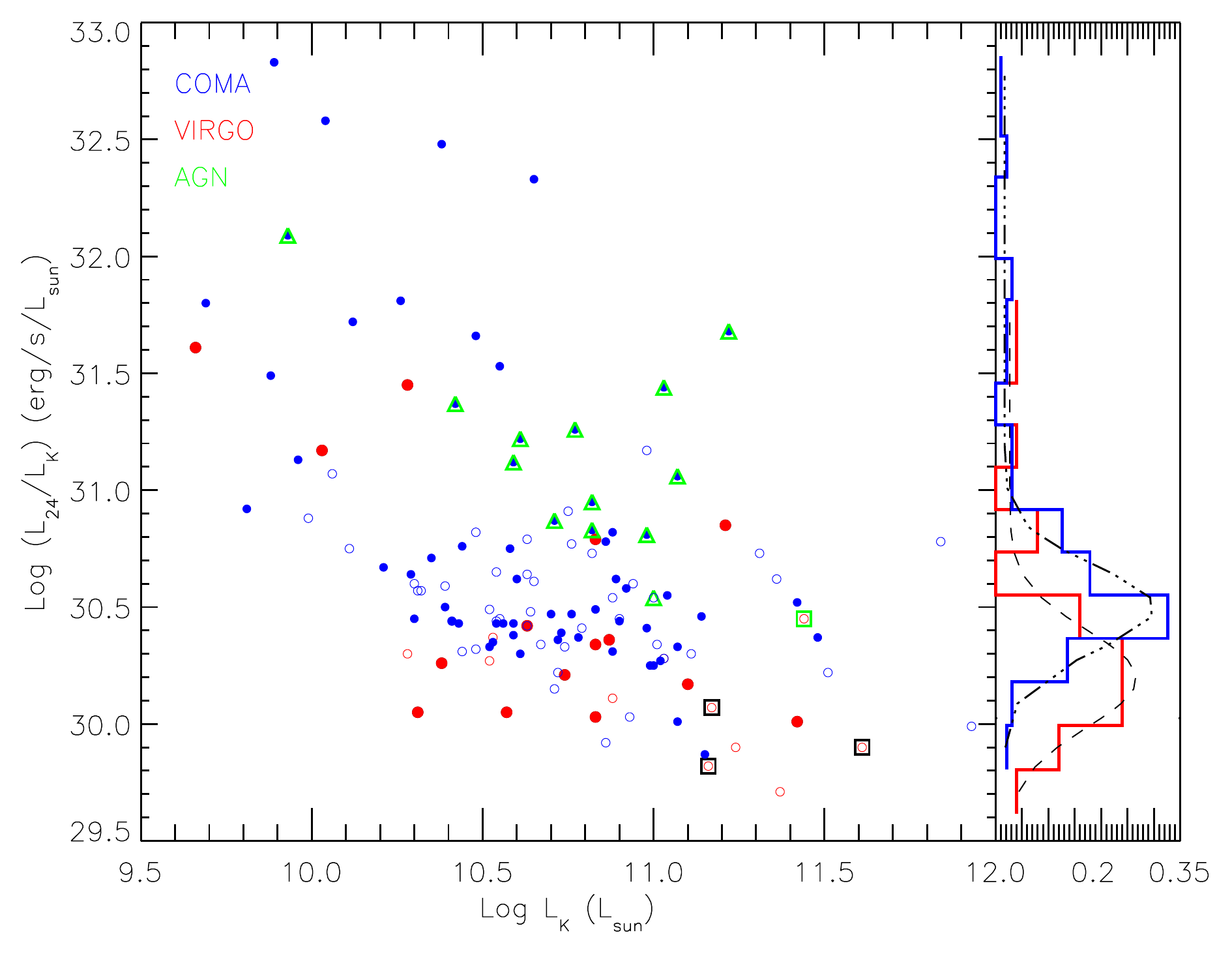}}
  \caption{L$_{24\mu m}$/L$_K$ distribution as a function of the K-band luminosity for our samples of early-type galaxies in Virgo (in red) and Coma (in blue). Elliptical galaxies are represented by open symbols while lenticulars are represented by filled ones. Known AGN sources are marked in green (green rectangles for Virgo and green triangles for Coma). The histograms on the right side show the L$_{24\mu m}$/L$_K$ distribution for non-AGN sources. The black lines (dashed for Virgo and dotted-dashed for Coma) correspond to the best fit of each distribution. The 3 black rectangles are sources with extreme 24$\mu$m fluxes according to \citet{Leipski:12} as detailed in Sect.\ref{sec:leipski}.}
 \label{fig:L24_LK_LK}
\end{figure*}

A tight correlation between the 24$\mu$m luminosity and the K-band luminosity is expected for ellipticals, as already described in the literature \citep[e.g.,][]{Temi:07a,Temi:09b,Young:09}. Both studies conclude that most of the 24$\mu$m emission in elliptical galaxies is circumstellar in origin. \citet{Temi:09b} found a larger scatter in the 24$\mu$m - K-band relation for S0 galaxies than for ellipticals. Some S0 galaxies from their sample have the same star-forming properties as ellipticals, i.e. almost no star-formation, but some S0 galaxies have a larger L$_{24\mu m}$/L$_K$ ratio, which is interpreted as the signature of a subpopulation of younger stars by \citet{Temi:09b}. \citet{Young:09}  noticed a population of galaxies with an even larger L$_{24\mu m}$/L$_K$ ratio than \citet{Temi:07} and found that CO-rich ETGs have 24$\mu$m flux densities up to 15 times larger that sources with a pure circumstellar 24$\mu$m emission. 

\subsection{Detection of a bimodality}

We explored the 24$\mu$m-K-band relation for ETGs in both Virgo (red) and Coma (blue) down to the same depth as illustrated in Fig.\,\ref{fig:L24_LK_LK}, which presents the  L$_{24\mu m}$/L$_K$ ratio as a function of the K-band luminosity. Open and filled circles represent Elliptical and  Lenticular galaxies respectively. 
Green rectangles and triangles indicate sources with strong AGN in the Virgo and Coma cluster respectively,  for the rest of the study they will be discarded from our sample.

The three black squares represent sources for which the 24$\mu$m flux is not safely known, and will be discussed in the following subsection. Fig.\,\ref{fig:L24_LK_LK} shows that most galaxies lie at a constant 
L$_{24\mu m}$/L$_K$ ratio of about 30, which is the expected ratio for galaxies with a gas-poor interstellar medium (ISM). Elliptical galaxies are located toward the bottom of the plot as expected \citep{Temi:09b}, while S0s span a range of $\sim$2 orders of magnitude in L$_{24\mu m}$/L$_K$. 
A few galaxies, mostly S0s, have a much larger flux ratio, some possibly due to the presence of a strong AGN but some due to a stronger star formation rate. 
Some of these galaxies with a large L$_{24\mu m}$/L$_K$ ratio and a weak AGN activity have a flux ratio comparable to late-type galaxies and will be investigate in the next sections.

The distribution of  the L$_{24\mu m}$/L$_K$ ratio for each cluster for non-AGN sources (right side panel of Fig.\,\ref{fig:L24_LK_LK})  shows that the ratio peaks at a value 0.30 $\pm$ 0.08 dex larger  
for Coma than Virgo and that Coma flux ratio distribution is broader than Virgo. Such differences in the flux ratio distribution are not expected between ETGs of two local clusters. 
The error on the offset between the two distributions has been derived from the errors on the peak location of Virgo and Coma. To estimate these errors, we generated 100 simulations of our Virgo and Coma catalog.\\
The distribution of Virgo simulations peaks on average at 30.14$\pm$0.07 (our sample peaks at 30.16). The distribution of Coma simulations peaks on average at 30.46$\pm$0.03 (our sample peaks at 30.46). These simulations allow us to confirm that our 25 source sample is still large enough to contraint the difference in the peak location with enough statistical significance, the smaller richness of Virgo cluster is therefore not a statistical issue.

Reasons for the L$_{24\mu m}$/L$_K$ ratio difference for the two clusters will be explored and tested in the next section. It implies that 
ETGs in Coma have either a larger amount of circumstellar dust per stellar mass or that 
they have an additional amount of cold ISM dust. Testing these galaxies with observations 
at longer wavelength could allow to disentangle these 2 scenarios. Another explanation 
could be that there is some photometric errors in one of these or both sets which produce 
either an overestimation of 24 $\mu$m fluxes for Coma galaxies or an underestimation 
of its K-band fluxes.

\subsection{Validity of the datasets}

The differences observed in the  L$_{24\mu m}$/L$_K$ distributions between Coma and
Virgo could come from photometric errors, especially since the photometry was calculated
by different teams for these 2 clusters.
In this subsection, we check the validity of these datasets by comparing them with 
the literature and with the WISE 22$\mu$m catalog.

\subsubsection{Comparison with \citet{Leipski:12}}
\label{sec:leipski}

\begin{figure}%[h]
 \centering
 \resizebox{1.\hsize}{!}{\includegraphics{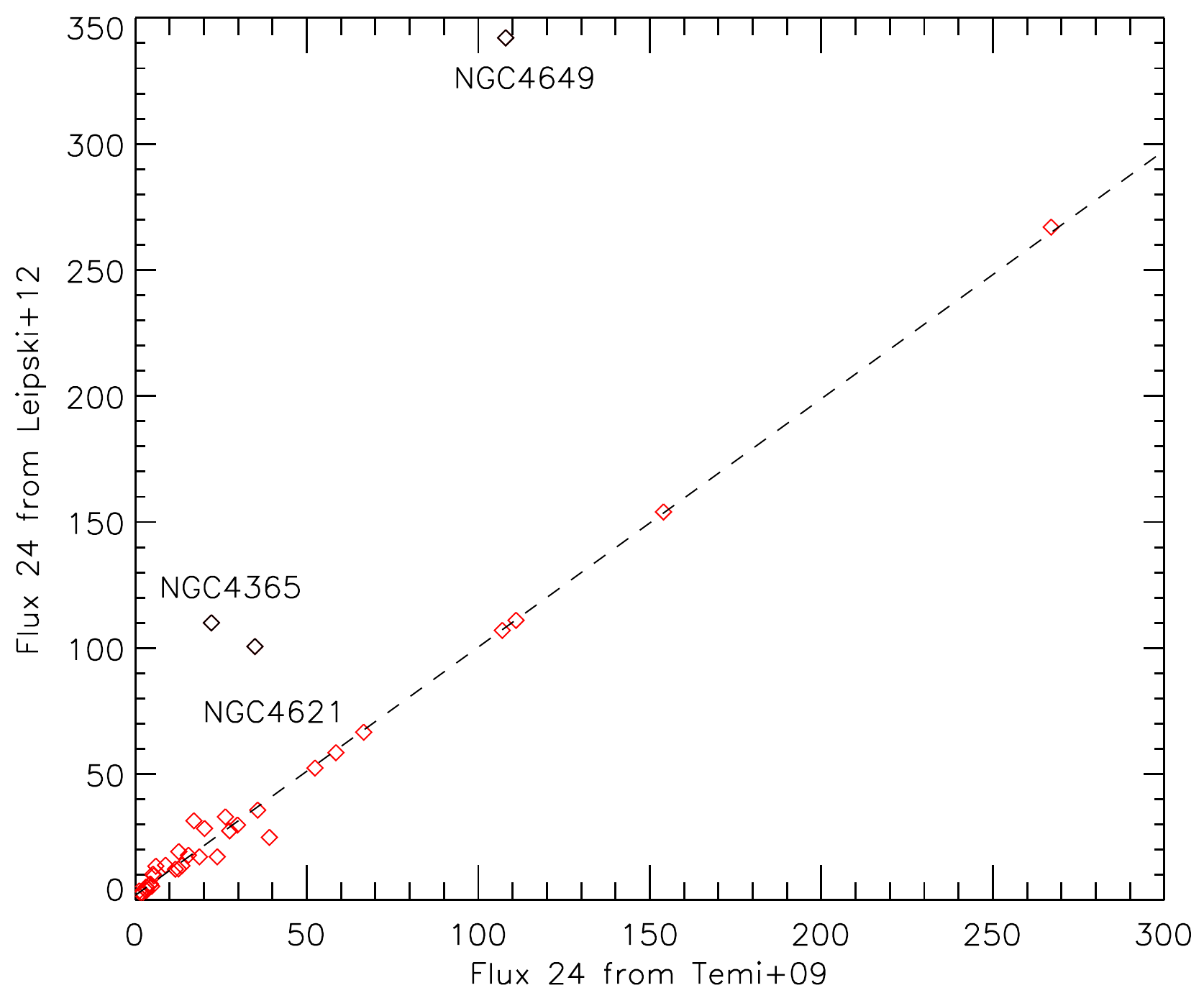}}
  \caption{Correlation between the 24$\mu$m flux estimates from \citep{Leipski:12} and \citep{Temi:09b}. The dotted line is a fit obtained while excluding 4 outliers (red diamonds with names, except for NGC4472, since \cite{Leipski:12} estimate is much larger) and the dot-dashed lines are 3-$\sigma$ limits. }
 \label{fig:correlation_f24}
\end{figure}

We find a  $\sim$0.5 dex difference between the L$_{24\mu m}$ distribution of Virgo and Coma with Virgo peaking at a shorter L$_{24\mu m}$ value, and a $\sim$0.25 dex difference for the L$_K$ distributions. 
Virgo cluster proximity (16\,Mpc) might make it hard to find the apertures that encapsulate the total flux of each galaxy. If the aperture is too small, fluxes might be underestimated. Furthermore larger apertures require to better estimate the background flux, since a small error in the background flux will have a huge impact on the estimated flux.
Since the major difference between Virgo and Coma is coming from the 24$\mu$m, we checked our 24$\mu$m fluxes \citep{Temi:09b} with values from the literature \citep{Leipski:12}. 
\citet{Temi:09b} and \citet{Leipski:12} have 51 sources in common, but only 43 of those are real detection at 24$\mu$m in \citet{Leipski:12}. The 8 other sources are only upper limits and were not used. Fig.\,\ref{fig:correlation_f24} shows a tight correlation between the 24$\mu$m fluxes from \citet{Temi:09b} on the x-axis and fluxes from \citet{Leipski:12} on the y-axis. 
Excluding 3 sources that show unrealistic flux estimates (NGC4365,  NGC4621, NGC4649), a 
linear fit to these data returns a  slope of (0.98$\pm$0.01) and a y-intercept of (1.85$\pm$0.79)\,mJy, showing that on average these flux estimates are identical.

These 3 outliers, marked with black squares on Fig.\,\ref{fig:L24_LK_LK}, are elliptical galaxies and their fluxes are not expected to be as high as estimated by \cite{Leipski:12}.
\citet{Bressan:07} looked at the IRS {\em Spitzer} spectroscopy of early-type galaxies. NGC4365 and NGC4621 are part of their sample and they found that these two galaxies show clear evidence for dust emission beyond 8$\mu$m which could explain the difficulties encountered to determine their 24$\mu$m fluxes. \citet{Martini:13} selected 38 ETGs to study their dust properties and they reported the 24$\mu$m fluxes of NGC4365, NGC4621, NGC4649 to be respectively: 31.2 mJy, 33.2 mJy and 59.4 mJy. These values are closer to the fluxes obtained by \citet{Temi:09b} (22.2+/-4.7 mJy, 34.9+/-6.3 mJy and 108+/-10 mJy respectively) than the ones obtained by \citet{Leipski:12} (110.01 mJy, 100.62 mJy and  342.10 mJy respectively).
This figure shows evidence that our 24$\mu$m fluxes are in agreement with another study and appear reliable.

\subsubsection{Confirmation with WISE data}

\begin{figure}%[h]
 \centering
 \resizebox{1.\hsize}{!}{\includegraphics{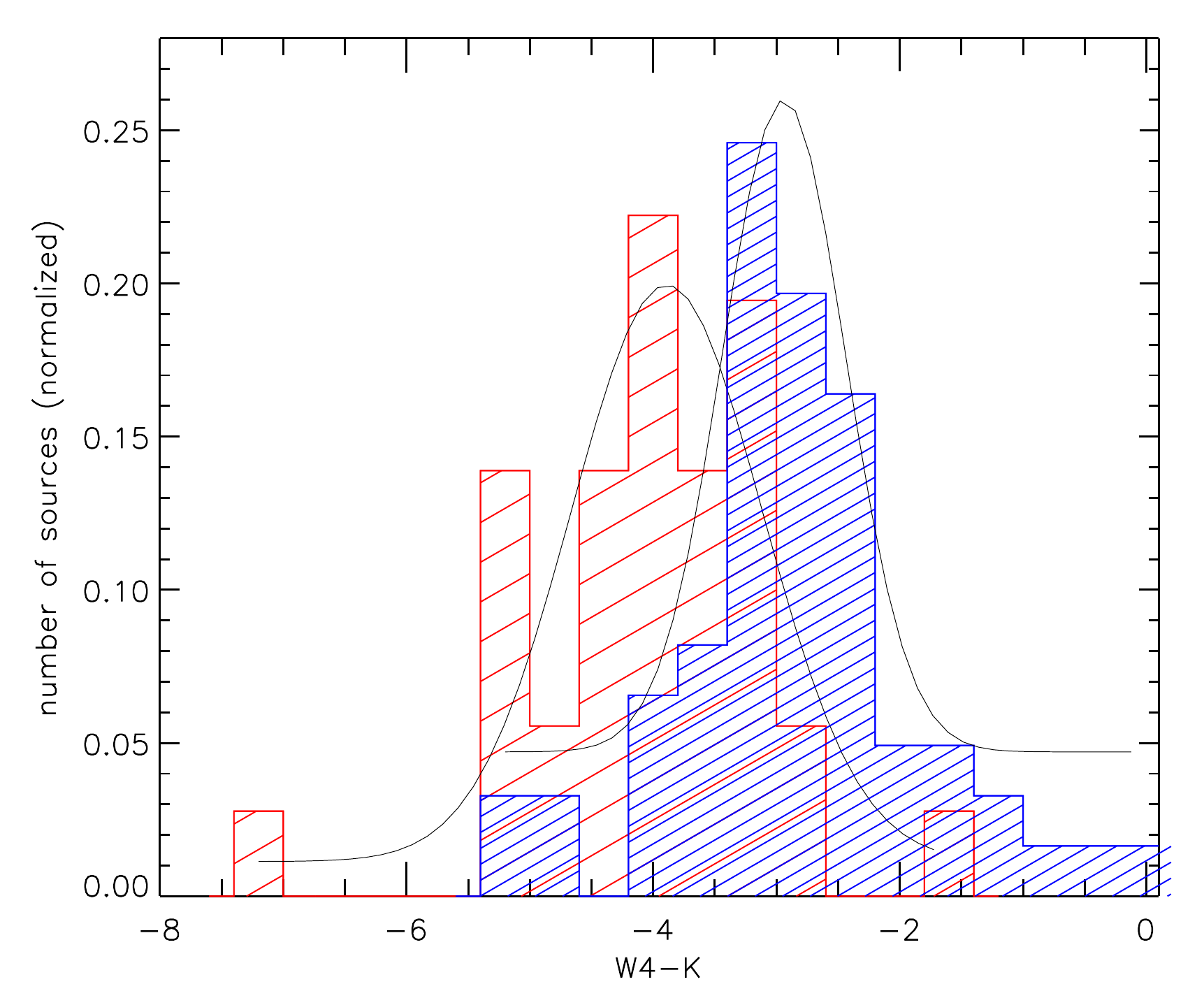}}
  \caption{W4-K distribution for Virgo (red) and Coma (blue) with the best fit over plotted in black. W4 is the 22$\mu$m channel of WISE. }
 \label{fig:WISE}
\end{figure}

Although WISE does not provide the same sensitivity as {\em Spitzer}, WISE has mapped the
 entire sky and its 22 $\mu$m band (W4) is similar in wavelength range to {\em Spitzer}/MIPS 24 $\mu$m filter.
 The AllWISE catalogue \citep{Wright:10,Mainzer:11} contains an aperture flux estimate for extended sources,
 {\it w4gmag}, that has been used for instance by \citet{Davis:14}. This catalog contains fluxes for both
 Virgo and Coma galaxies, that have been produced in the same fashion and should be less prone to systematic 
difference between the two clusters.
25 of our Virgo ETGs and 63 of our Coma ETGs are detected at 22$\mu$m by WISE down to 3.6\,mJy, which corresponds to a magnitude of 8.4 [Vega] for Coma and 4.5 [Vega] for Virgo.
 Fig.\,\ref{fig:WISE} shows the distribution of the W4-K color for Virgo in red and Coma in blue.
The two distributions are shifted with respect to each other by less than 1 dex in color,
i.e. a $\sim$0.37\,dex difference in luminosity, with Coma presenting a larger MIR/K ratio, as found previously
with MIPS-24$\mu$m. This reinforce our findings that Coma ETGs have, on average, a larger MIR-to-K luminosity ratio.
A similar result is found even if we limit our analysis to the distributions of elliptical galaxies in the two clusters: a difference of $\sim 0.35$ dex in the mean value of the two distribution is measured for ellipticals.\\

\subsection{Tentative physical interpretations}

Such a discrepancy can be ascribed to differences in the local environment, and the physical and dynamical conditions of the two clusters, as they have a profound effect on the galaxy population, the evolution and transformation of galaxies
 and their star formation history.

Several channels of morphological transformation of disk galaxies into ETGs may be at play in cluster environment, including ram-pressure stripping of cold gas through interaction with the dense cluster medium \citep{Abadi:99, Quilis:00, Smith:10}, harassment \citep{Farouki:81,Moore:99} and gas consumption by star formation \citep{Larson:80, Bekki:02}. The interpretation of our data regarding the Coma cluster suggests that {\it pre-processing}
 due to environmental processes in small groups plays an important role in galaxy transformation and star formation activity before they enter the denser cluster environment.
In a subsequent paper, we will investigate possible links between the morphological transformation and physical processes responsible for quenching the star formation in Virgo and Coma clusters. The discrepancy in the L$_{24\mu m}$/L$_K$
 ratio of the ETG distributions in the two clusters must be explained in relation to the physical properties of their environments.
Here we focus our attention on a sub-population of ETGs in Coma and Virgo by studying their star formation activity and by looking for radial trends and possible associations with infalling galaxy groups within the cluster.
Recent studies of (optically) red-sequence galaxies in Coma \citep{Smith:12, Rawle:13} have shown that the transformation of spirals into S0s is associated with cluster infall \citep{Rawle:13} and that spatial trends in
 the ages of these galaxies are observed within the cluster \citep{Smith:12, Rawle:13}.

\section{Mid-Infrared Enhanced Galaxies (MIEGs): extreme sources in the Coma \& Virgo clusters}
\label{MIEG}

\subsection{The Mid-Infrared Enhanced Galaxies (MIEGs) selection }

\begin{figure}%[htpb]
\resizebox{1.2\hsize}{!}{\includegraphics{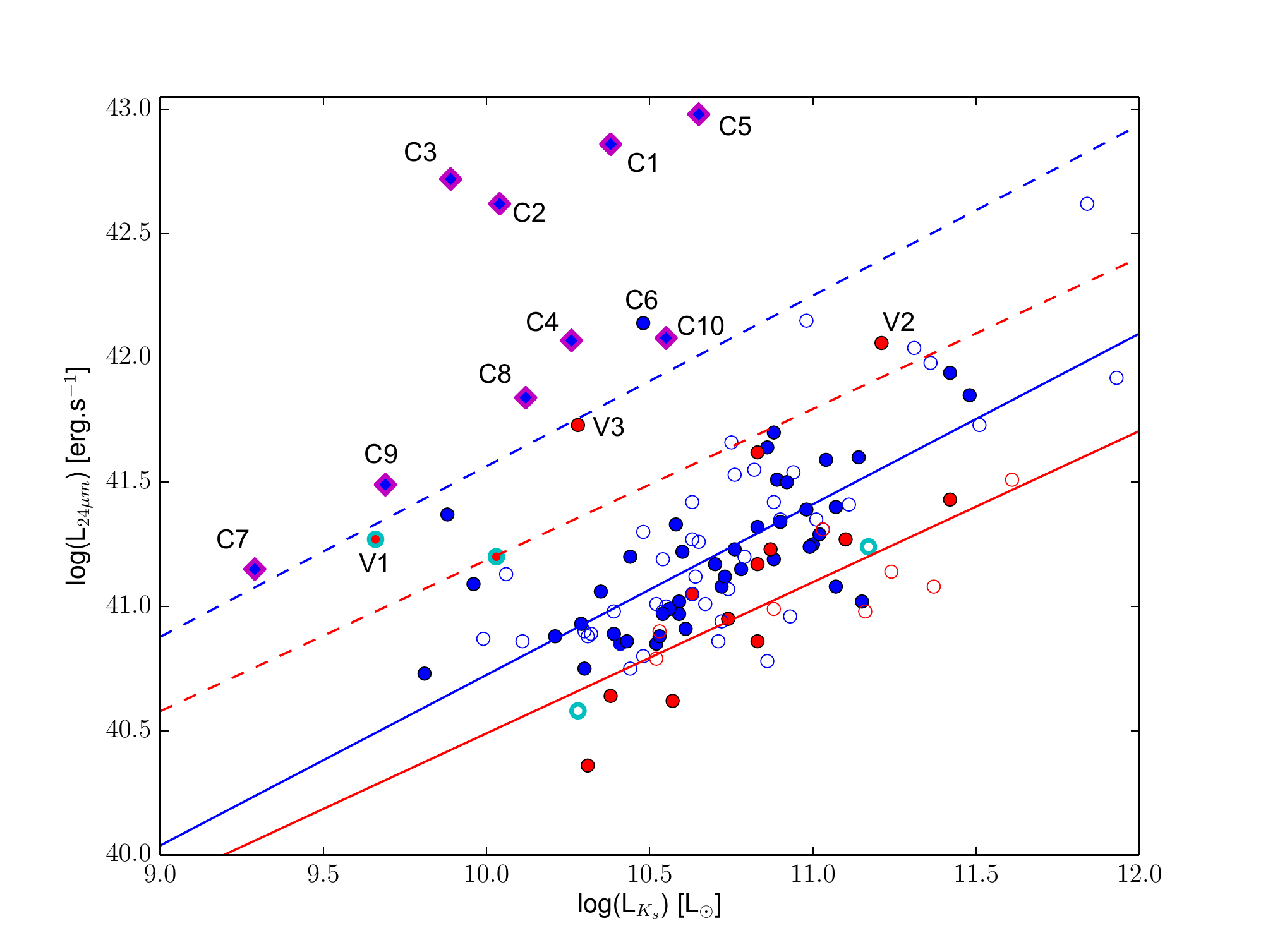}}
  \caption{24$\mu$m luminosity as a function of the K-band luminosity in logarithmic scale for early-type galaxies in the Coma (in blue) and Virgo (in red) clusters. Magenta(cyan) diamonds represent star-forming sources in Coma(Virgo) as inferred with the BPT diagram diagnostic. The blue and red dashed lines represent the MIEG selection line for Coma and Virgo galaxies respectively, the solid lines represent the fitted log(L$_{24\mu m}$)-log(L$_K$) relation.}
 \label{fig:L24_LK_MIEGs}
\end{figure}

\begin{deluxetable*}{lccccccc}
\tablecaption{Coma \& Virgo Mid-Infrared Enhanced Galaxies
\label{table:t_param}
}
\tablewidth{0pt}

\tablehead{
Galaxy & Name & RA & Dec & cz & $r$ & $M_{r}$ & $(g-r)$\\
\colhead{}  & \colhead{}  &  \colhead{}  &  \colhead{}  & 
\colhead{km sec$^{-1}$}  & \colhead{(mag)}  &  \colhead{(mag)}  & \colhead{(mag)}\\
\colhead{(1)}  &  \colhead{(2)}   &  \colhead{(3)}   &  \colhead{(4)} & 
\colhead{(5)}  &  \colhead{(6)}   &  \colhead{(7)}   &  \colhead{(8)}\\
 }
 
\startdata
C1   & Mrk 57                          & 194.6554  & 27.1762  & 7647  & 15.00 & $-$20.24 & 0.36\\
C2   & Mrk 56                          & 194.6472  & 27.2649  & 7351  & 15.23 & $-$19.92 & 0.28\\
C3   & Mrk 53                          & 194.0254  & 27.6781  & 4951  & 15.05 & $-$19.23 & 0.34\\
C4   & J12570456+2746228   & 194.2689  & 27.7730  & 7528  & 15.66 & $-$19.55 & 0.66\\
C5   & NGC 4926A                  & 195.5328  & 27.6483  & 6889  & 14.36 & $-$20.65 & 0.57\\
C6   & J12581382+2810576   & 194.5575  & 28.1825  & 7169  & 14.71 & $-$20.39 & 0.68\\
C7   & J12575620+2734524   & 194.4843  & 27.5814  & 4943  & 16.08 & $-$18.20 & 0.14\\
C8   & J12583788+2727497   & 194.6577  & 27.4641  & 6287  & 15.69 & $-$19.12 & 0.58\\
C9   & J12582052+2725457   & 194.5855  & 27.4294  & 7562  & 16.78 & $-$18.43 & 0.59\\
C10 & J12584321+2854356   & 194.6800  & 28.9101  & 8353  & 14.96 & $-$20.47 & 0.57
\vspace{0.05cm}\\
\hline
\vspace{-0.15cm}\\
V1   &  NGC 4344   &  185.906  & 17.541   & 1058  &  12.46  &  $-$18.80  &  0.65\\
V2   &  NGC 4526   &  188.513  & 7.699  & 494  &  9.73  &  $-$21.41  &  0.86\\
V3   &  NGC 4694   &  192.063  & 10.984  & 1061  &  11.55  &  $-$19.75  &  0.41\\

\enddata
\tablecomments{Col. (1): Designation.  
Col. (2): Source name.  
Col. (3) \& (4): J2000 epoch right ascension and declination as listed in NED.   
Col. (5): Recessional velocity, corrected to the local group centroid, as reported in NED. 
Col. (6): SDSS r-band Petrosian magnitude taken from the data release 8 and corrected for foreground extinction. 
Col. (7): Absolute r-band magnitude using the distances listed in Table 3 and 4. 
Col. (8): ($g-r$) color, corrected for foreground extinction.  }
\end{deluxetable*}

\citet{Temi:09a,Temi:09b} have shown that the K-band luminosity of elliptical galaxies correlates tightly with L$_{24\mu m}$. This correlation is approximately log\,L$_{24\mu m}$\,$\simeq$\,log\,L$_K$\,+30.5, where L$_{24\mu m}$ is expressed in erg/s and L$_K$ in solar luminosity. Such a tight correlation is expected among elliptical galaxies since the 8-24$\mu$m flux is produced by the collective emission from circumstellar dust flowing away from red giant stars. This mid-infrared emission has a de Vaucouleurs profile similar to optical starlight \citep{Athey:02,Temi:08}. When the log(L$_{24\mu m}$/L$_K$) exceed the threshold of 30.5 it becomes a robust measurement of the specific star formation.
We will focus our work on the most extreme sources, i.e. with the largest L$_{24}$/L$_K$ ratio, in Fig.\,\ref{fig:L24_LK_LK}. They happen to be mostly lenticulars (filled symbols). Another way to look at the L$_{24}$/L$_K$ relation is to plot the L$_{24}$ luminosity as a function of L$_K$ as illustrated on Fig.\,\ref{fig:L24_LK_MIEGs}.
We ran RANSAC (random sample consensus), an iterative fit method, on the Coma and Virgo data points to estimate a linear fit between the 24$\mu m$ and K band fluxes.
After 50 iterative steps, the mean fitted relations are log(L$_{24\mu m}$) =  0.69$\times$ log(L$_K$) + 33.8 for Coma and log(L$_{24\mu m}$) =  0.61$\times$ log(L$_K$)+ 34.4 for Virgo (solid lines on Fig.\,\ref{fig:L24_LK_LK}).

With the goal to select sources lying well above the K-24$\mu$m correlation, we select only sources above the upper 5-\,$\sigma$ line, 
\begin{eqnarray}
log\,L_{24\mu m} &> & 0.69\times log\,L_K\,+\,34.7 \mathrm{\;\;for\; Coma},\nonumber\\
log\,L_{24\mu m} &> & 0.61\times log\,L_K\,+\,35.1 \mathrm{\;\;for\; Virgo}
\end{eqnarray} 
These sources are defined as Mid-Infrared Enhanced Galaxies (MIEGs) and we
will focus on these sources for the rest of this work. The Virgo MIEG selection
line lies below Coma's one because its galaxies have a lower \rattk\, on average, as described in the previous section. 
10 Coma MIEG sources are lying above the Coma selection line (from C1 to C10), and 3 Virgo MIEG sources are lying above the Virgo selection line (from V1 to V3).
These 13 MIEGs are all lenticulars.
Fig.\,\ref{fig:L24_LK_MIEGs} shows the L$_{24}$ distribution as a 
function of L$_K$ for both Virgo (red) and
 Coma (blue) sources. The MIEG selection cut-offs (Eq.\,1) are represented by  dashed blue and red lines. 
\citet{Davis:14} fitted a similar relation between WISE 22 $\mu$m and 2MASS K-band for ATLAS$^{3D}$ ETGs 
that do not have a detection of CO molecular gas. Their best-fit (log(L$_{22\mu m}$) =  1.00$\times$ log(L$_K$) + 30.5) is in good agreement with the L$_{24}$-L$_K$ relation for the bulk of the Coma and Virgo ETG population. 
Their results will be compared to our work in details in Sect.\,\ref{sec:discussion}. 
The BPT diagnostic gives us informations on the nature of the ETG population: 
pink diamond (Coma) and light blue circles (Virgo) mark known star-forming sources. 
The large L$_{24\mu m}$/L$_K$ value of these extreme sources can not be explained 
by AGN activity as 10 sources among the 13 MIEGs are known to be dominated by star-formation.

\subsection{Morphology of the Coma MIEGs}
\label{sec:morph}

\begin{figure*}
% \centering
  \resizebox{1.\hsize}{!}{\includegraphics{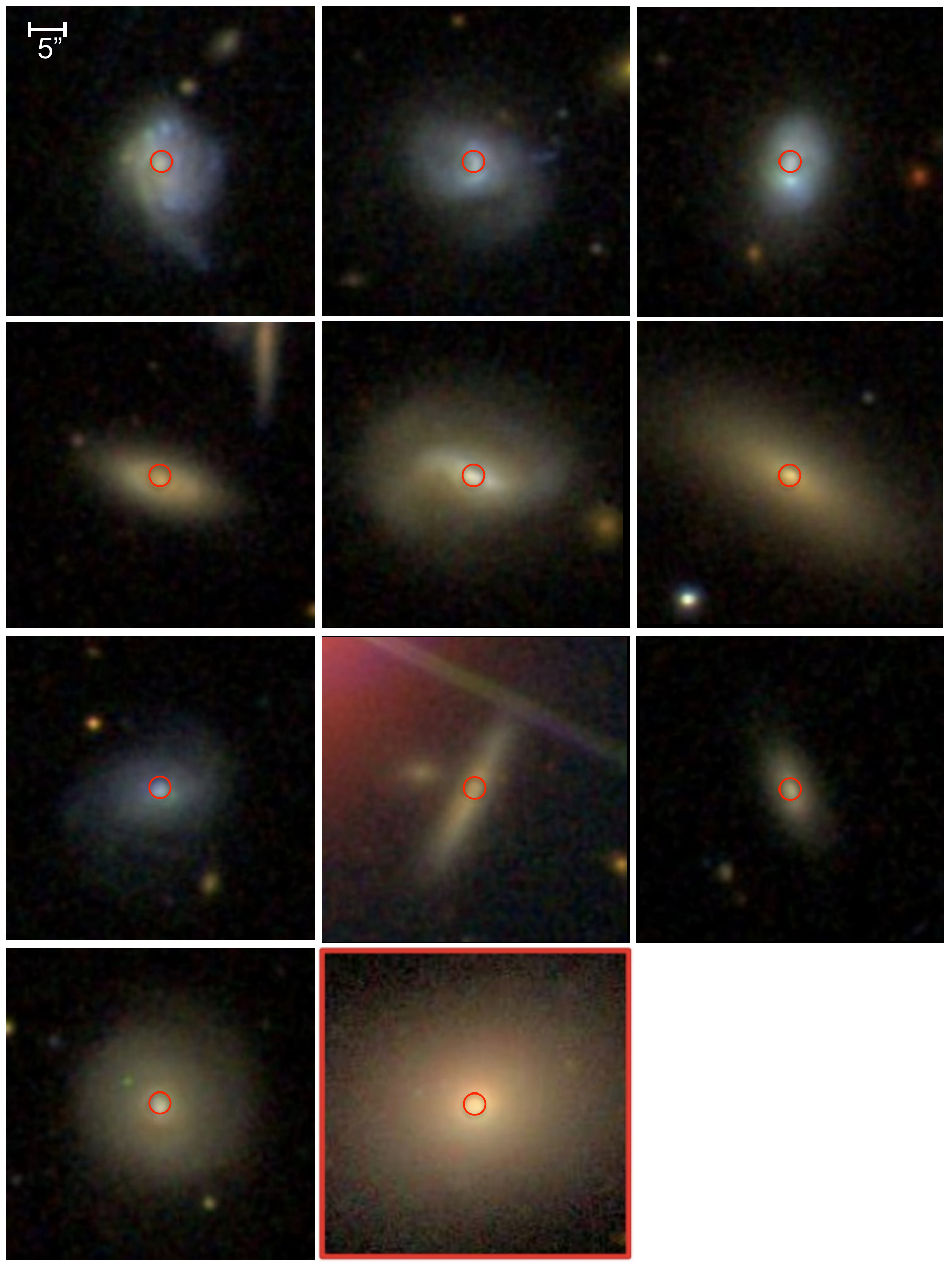}}
  \caption{40'' $\times$ 40'' SDSS images of the MIEGs: first row from left to right: C1, C2, C3; second row from left to right, C4, C5, C6; third row from left to right: C7, C8, C9; fourth row from left to right: C10 and a typical ETG in Coma, framed in red. The SDSS fiber size (3'' diameter) is represented at the center of the images as a red circle.}
 \label{fig:SDSS}
\end{figure*}

\begin{figure*}
 \centering
  \resizebox{0.33\hsize}{!}{\includegraphics{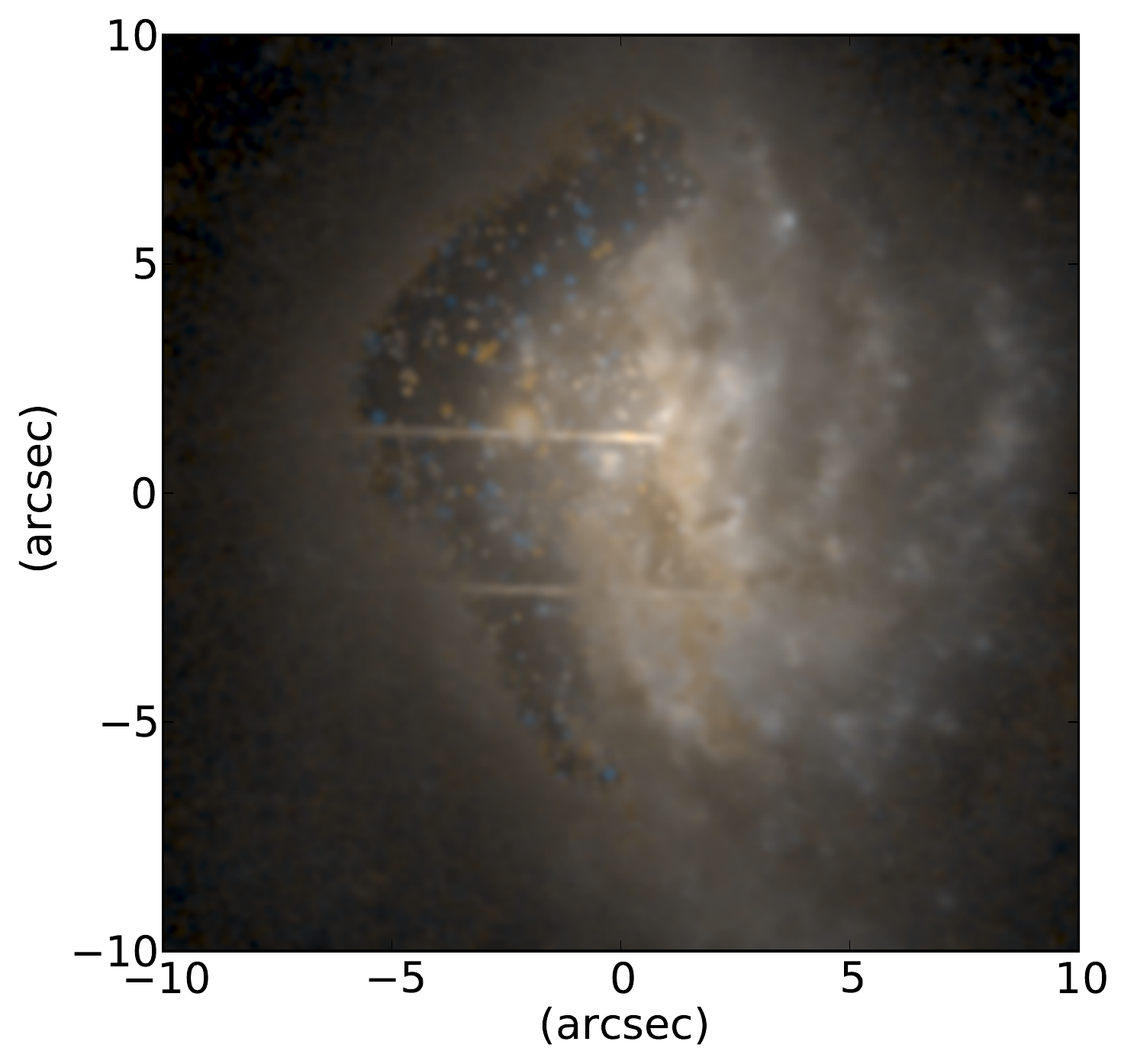}}
  \resizebox{0.33\hsize}{!}{\includegraphics{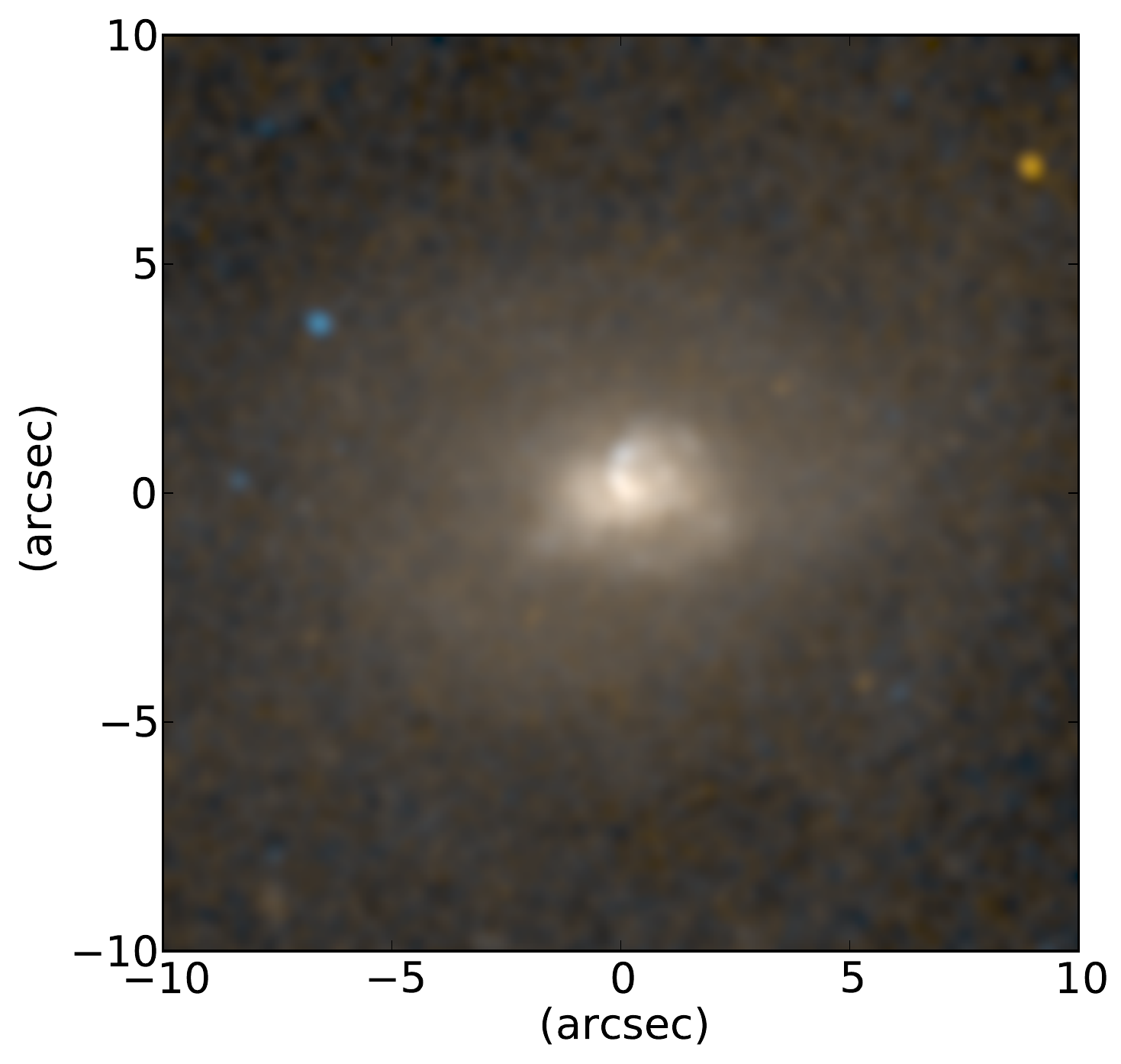}}
  \resizebox{0.33\hsize}{!}{\includegraphics{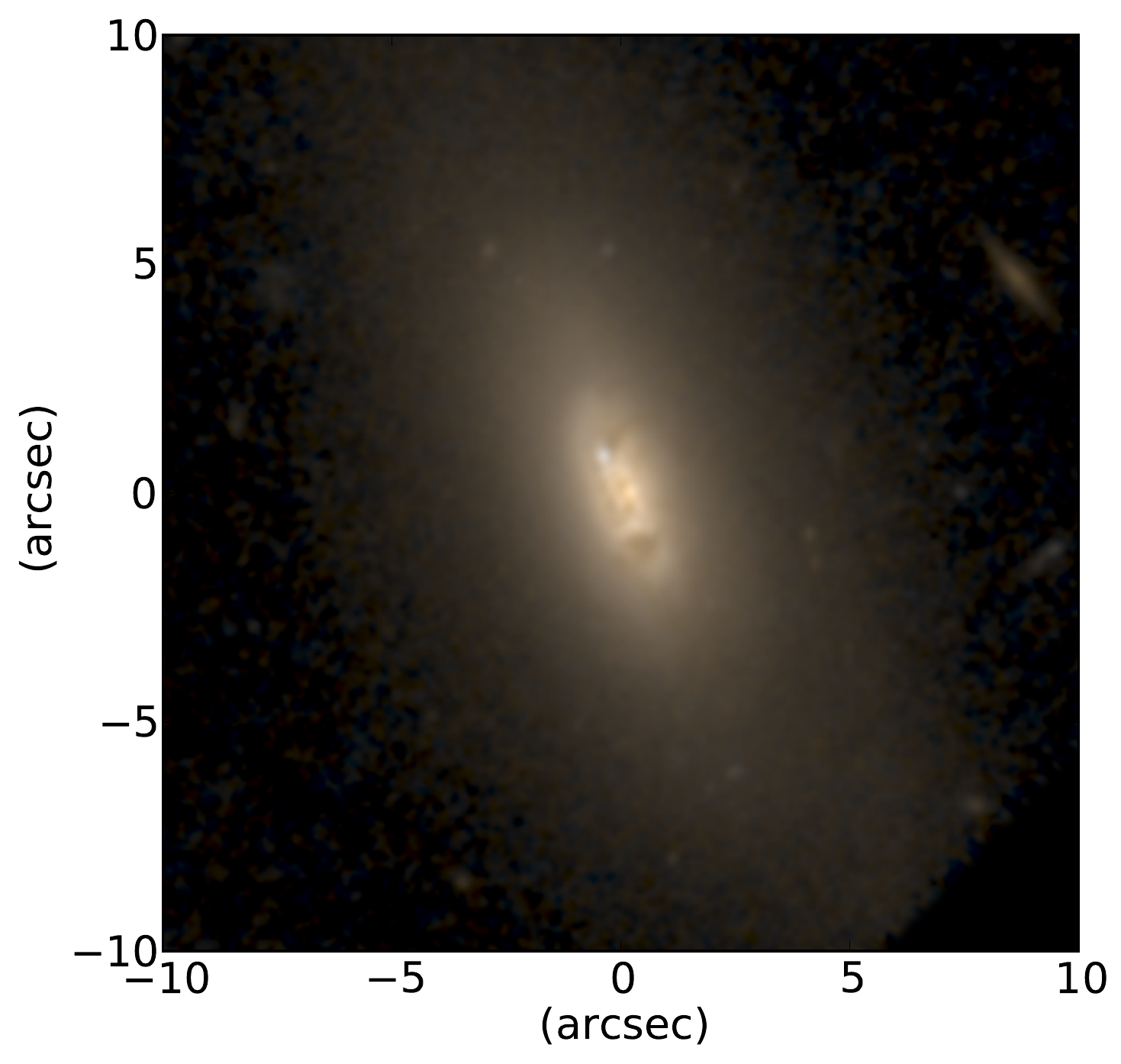}}
  \caption{RGB images, produced with 2 HST filters,  F814W and F450W, for some MIEGs: C1 (left), C7 (middle), C9 (right). The contrast of the intensity has been reduced by taking its value to the power of 0.4. All these MIEGs have a disturbed surface brightness, atypical for ETGs, with some blue knots and dust absorption structures.}
 \label{fig:HST}
\end{figure*}

We focus on the 10 Coma galaxies lying above the 2-$\sigma$ line displayed in Fig.\,\ref{fig:L24_LK_MIEGs}; these sources are labeled C1 through C10.  Table\,\ref{table:t_param} provides additional information on these 10 sources.   
To understand their morphological and photometric properties, we examined imagery from the SDSS and other archives.  Three of the ten have been imaged with the Hubble Space Telescope (HST), permitting a higher resolution inspection.  In Fig.\,\ref{fig:SDSS} we present SDSS 3-color images of all 10; in Fig.\,\ref{fig:HST}, HST/ACS and HST/WFPC2 images in the F450W and F814W filters are used to produce RGB images of C1, C7 \& C9.  

\subsubsection{  C1 - Mrk 57}

C1 presents the most uncertain classification in this set. The SDSS image shows a disturbed blue disk with a likely tidal feature to the south.  We note that Mrk\,56, C2 in our list, is located $\sim$5' to the north of C1, corresponding to a linear separation of $\sim$150 kpc at Coma with a $\Delta$V $\sim$ 300 km / sec. Both are distorted and actively forming stars.  In the higher resolution HST image, C1 presents a more irregular morphology, with a prominent dust lane and blue knots.  These images do not permit a clean classification for C1, which might be a disturbed late-type system or a gas-rich S0 undergoing a mild, interaction-induced starburst.  We decide to keep C1 as part of our MIEG sample, but with a caution.

\subsubsection{  C2 - Mrk 56}

Similar in global color and luminosity to C1 \& C3, C2 shows a distorted, diffuse, largely face-on disk, with a shell or tidal loop to the SW.  The core appears elongated N-S with little evidence of a redder bulge.  No substructure is obvious in the SDSS image.  

\subsubsection{  C3 - Mrk 53}

The disk in C3 displays smoother outer isophotes compared to C1 \& C2 with no prominent distortions and appears typical of "blue cloud" ETGs.  In the center, there is a single, blue knot, offset $\sim$3" S from the redder bulge and the hint of a dust patch to the west.  

\subsubsection{  C4}

C4 shows an oval disk, a red bulge and globally smooth isophotes.  A circumnuclear ring is likely present which is the source of the 24$\,\mu m$ emission.  

\subsubsection{  C5 - NGC 4926A}

C5, the most optically luminous of the candidate MIEGs, contains a pear-shaped, diffuse outer disk in which a stellar bar extends from an elongated nuclear bulge at a PA $\sim$70$^{\circ}$. The bar shows  a distinct twist with radius. This system is also the brightest 24$\mu$m source.    

\subsubsection{   C6}

Similar to C4, this system displays smooth outer isophotes, an inclined, generally red disk with no structures evident in the SDSS image.   The bulge is more prominent compared to C4, consistent with the 1.2 mag difference in absolute r magnitude between C4 \& C6.  

\subsubsection{  C7}

C7 has the oddest morphology in this sample. It is the bluest object, in (g-r), similar to C1-3, but about a factor of two less luminous.  A prominent blue core is present within a somewhat irregular outer disk.  The blue core and distorted disk are consistent with a global starburst, concentrated in the nucleus.  The HST/WFPC2 image shows two intense blue knots in the core, each a few hundred parsecs across, embedded in an irregular patchy region about a kpc in diameter, which may be a circumnuclear ring. No spiral arcs or similar features are evident, suggesting that C7 is an ETG, despite the large formal uncertainty on its assigned T value. \citet{Caldwell:93,Caldwell:99} included C7 in a list of early type galaxies with enhanced star formation in the Coma infall region, designating it as an S0 but also quoting the Sa classification from \cite{Dressler:80}.  

\subsubsection{ C8}
C8 is a highly inclined disk system with a clear radial color gradient and no prominent bulge. Due to its observed orientation, a robust classification is challenging. The absence of blue knots or a central dust lane suggests that C8 is properly an S0+ \citep[e.g.,][]{Kormendy.Bender:12}.

\subsubsection{  C9}

C9 exhibits a small, diffuse disk with a modest central bulge.  It is the least luminous, in SDSS r, of the candidate MIEGs and is intermediate in (g-r) color between the blue systems and the redder C6 \& C8.  The HST/ACS image shows structure in the inner kpc, possibly a ring or dust lane; at larger radii the disk is azimuthally symmetric, validating the ETG designation.  

\subsubsection{  C10}

C10 appears as a face-on S0 with a bright core and a smooth, slightly asymmetric disk. The central region is bluer than typical SO bulges with a dust patch visible between PA 160\degree and 190\degree. The disk shows a lower surface brightness extension in the NW quadrant. Globally, C10 is a slightly less luminous and more symmetric version of C5.

\subsection{ Summary}

The candidate MIEGs presented here exhibit diverse morphologies, ranging from globally blue, asymmetric systems to redder, more relaxed galaxies.  Seven show resolved evidence for  enhanced star formation in addition to the 24$\mu$m flux. C6 \& C8 appear as normal, quiescent early-type galaxies, although we note that C8 (but not C6) is a GALEX FUV source.  We note that the observed redshifts for C3 and C7 are somewhat lower than the Coma cluster mean redshift, but within the range for possible current and future cluster members, given the large velocity dispersion of the cluster. These two systems may be part of the structures currently infalling into Coma, from either the foreground or background.  In section \ref{sec:analysis}, we examine the location of these MIEGs within the Coma region to explore the origin of their enhanced mid-IR emission with respect to the cluster substructure and thermodynamics of the hot gas.

\subsection{Morphology of the Virgo MIEGs}
\label{sec:vmorph}

\begin{figure*}
 \centering
  \resizebox{0.7\hsize}{!}{\includegraphics{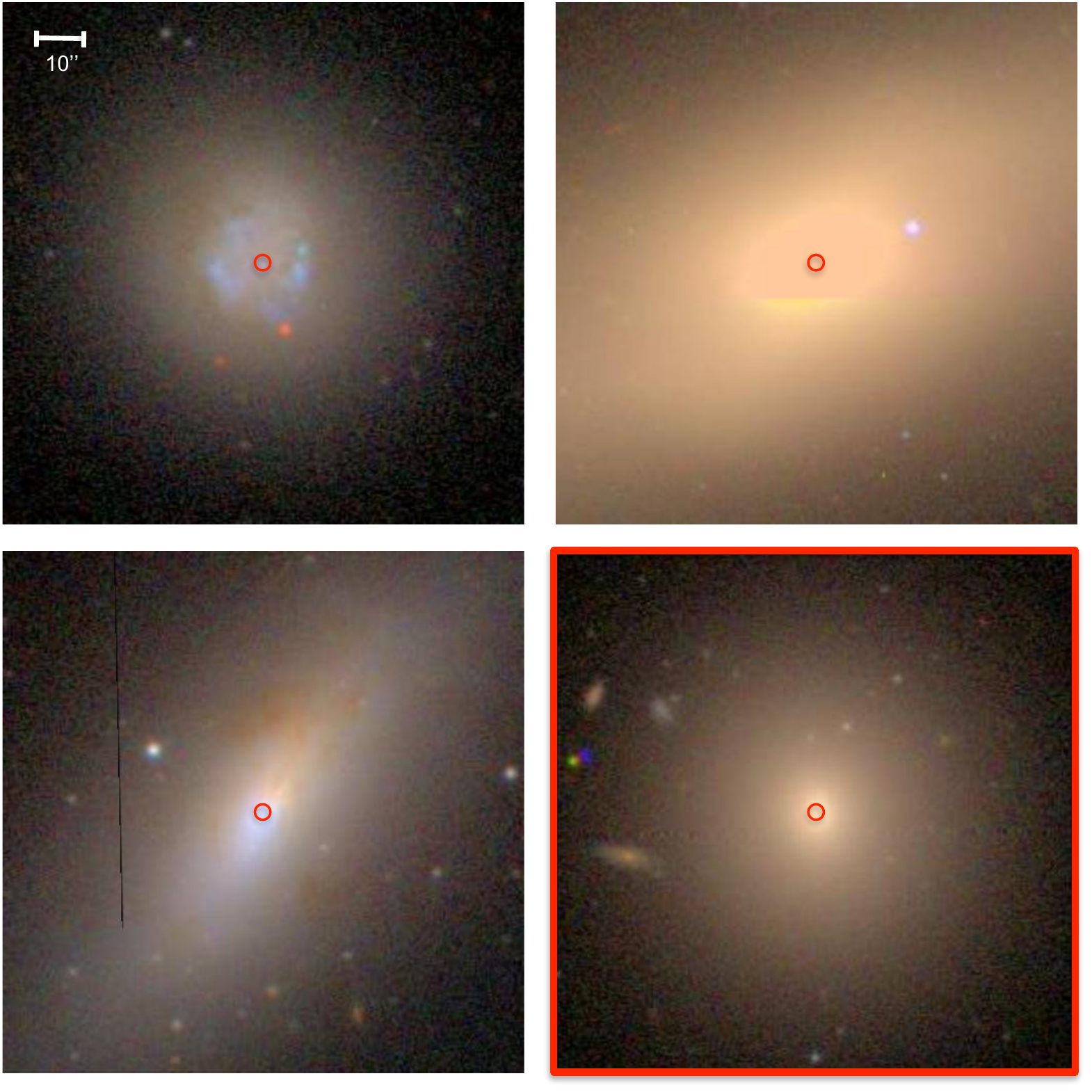}}
  \caption{80'' $\times$ 80'' SDSS images of the MIEGs: first row from left to right: V1, V2; second row from left to right, V3 and a typical ETG in Virgo (framed in red). The SDSS fiber size (3'' diameter) is represented at the center of the images as a red circle.} 
 \label{fig:SDSS_v}
\end{figure*}

\begin{figure}
 \centering
  \resizebox{1.\hsize}{!}{\includegraphics{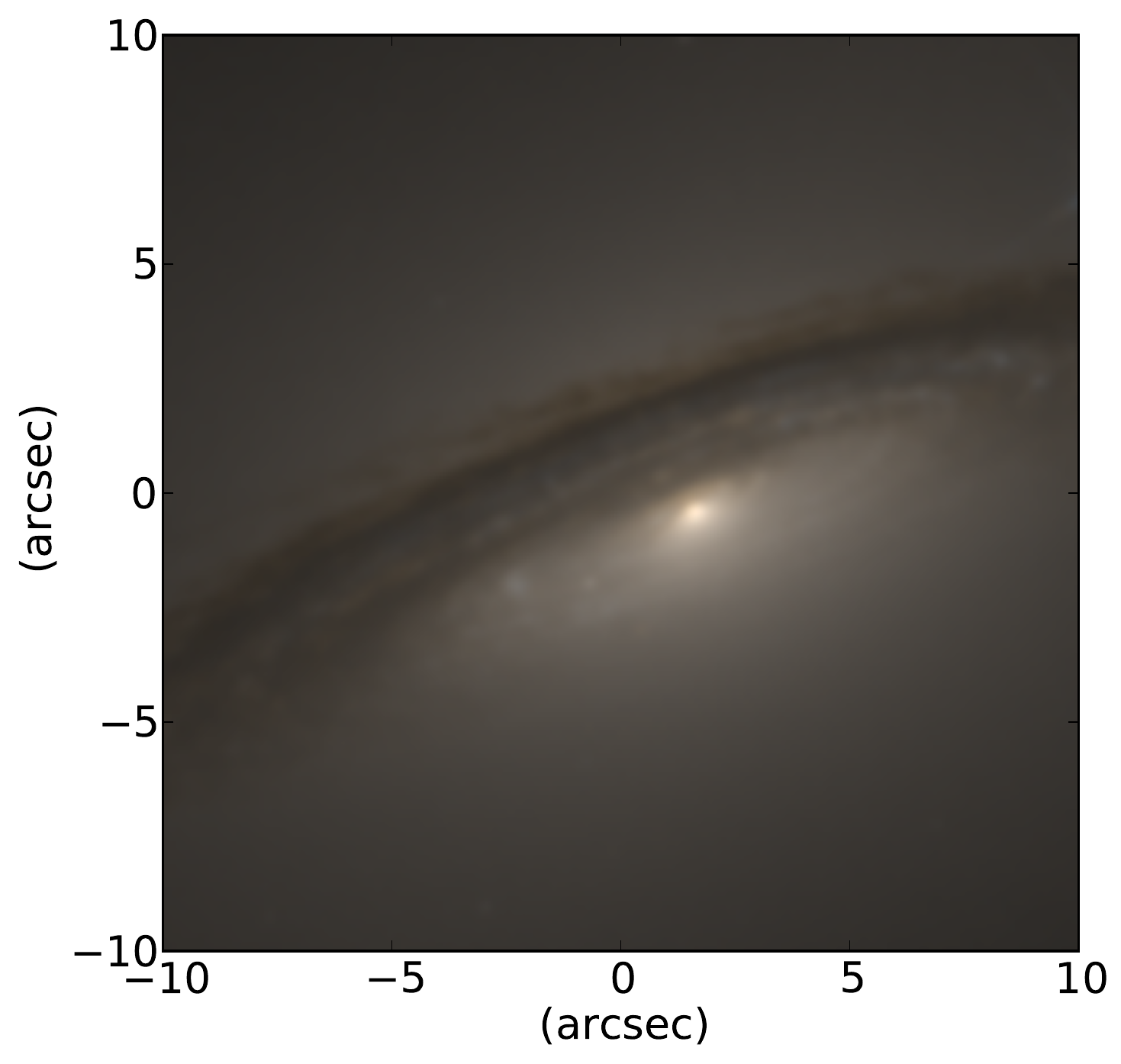}}
  \caption{RGB image for the Virgo-MIEG V2, produced with 2 HST filters, among  F475W, F555W, F850LP and F814W. The contrast of the intensity has been reduced by taking its value to the power of 0.4.}
 \label{fig:HST_v}
\end{figure}

We focus on the 3 objects lying above or close to the 2-$\sigma$ line displayed in Fig.\,\ref{fig:L24_LK_MIEGs}; these sources are labeled V1 through V3.  Table\,\ref{table:t_param} provides additional information on these 5 sources. To understand their morphological and photometric properties, we examined imagery from the Hubble, SDSS and other archives. 
In Fig.\,\ref{fig:SDSS_v} we present SDSS 3-color images of all 3. 

\subsubsection{  V1 - NGC4344} 
NGC 4344 is a face-on S0 whose optical appearance is dominated by a prominent, 
broken ring of blue starburst knots.  Patches of dust are visible also, possibly 
causing the gaps in the ring.  At the adopted distance, this ring has a diameter
 of $~1.7$ kpc.  No HST images are available for this system.  

\subsubsection{  V2 - NGC4526} 

The SDSS images show that NGC 4526 is an inclined S0 (i $\sim$ 78$\degree$) with a smooth outer disk and a prominent bulge surrounded by a substantial dust disk.  This structure is well-resolved in Hubble ACS images (Fig.\,\ref{fig:HST_v} which show a complex, filamentary dust distribution, visible at all angles around the bulge.  Both the multicolor Hubble images and integral field spectra from SAURON (Kuntschner:06) indicate the presence of ongoing high-mass star formation with this dust disk.  Reprocessed UV light from these young stars powers the enhanced 24$\mu$m emission.

\subsubsection{  V3 - NGC4694}

NGC 4694 is a highly-inclined S0 with a blue core and asymmetric dust 
patches visible in the SDSS images.  The Hubble data shows a chaotic
dust morphology, more similar to that seen in VCC 571 than the more 
symmetric disk-like feature seen in NGC 4476 and NGC 4344. Outer 
isophotes are smooth, providing little evidence that the stellar disk has been perturbed.

\subsection{ Conclusions on the Virgo -MIEGs }

Blue knots are
interspersed within each structure indicating that young stars are forming within these disks.  
We can conclude that the MIEG phenomenon in 
the Virgo systems arises from an excess amount of dust, which appears to be related to the stellar 
disk most but not all systems.  

Mapping of the location of the Virgo MIEGs within the cluster shows no correlation with position 
or the ROSAT-derived hot gas distribution.  This
result is different than that found for Coma, suggesting that more than one process can enhance 
the dust emission in early-type galaxies.

\section{Analysis}
\label{sec:analysis}

\subsection{Distribution of the MIEGs among the cluster}

\begin{figure*}%[h]
 \centering
  \resizebox{0.8\hsize}{!}{\includegraphics{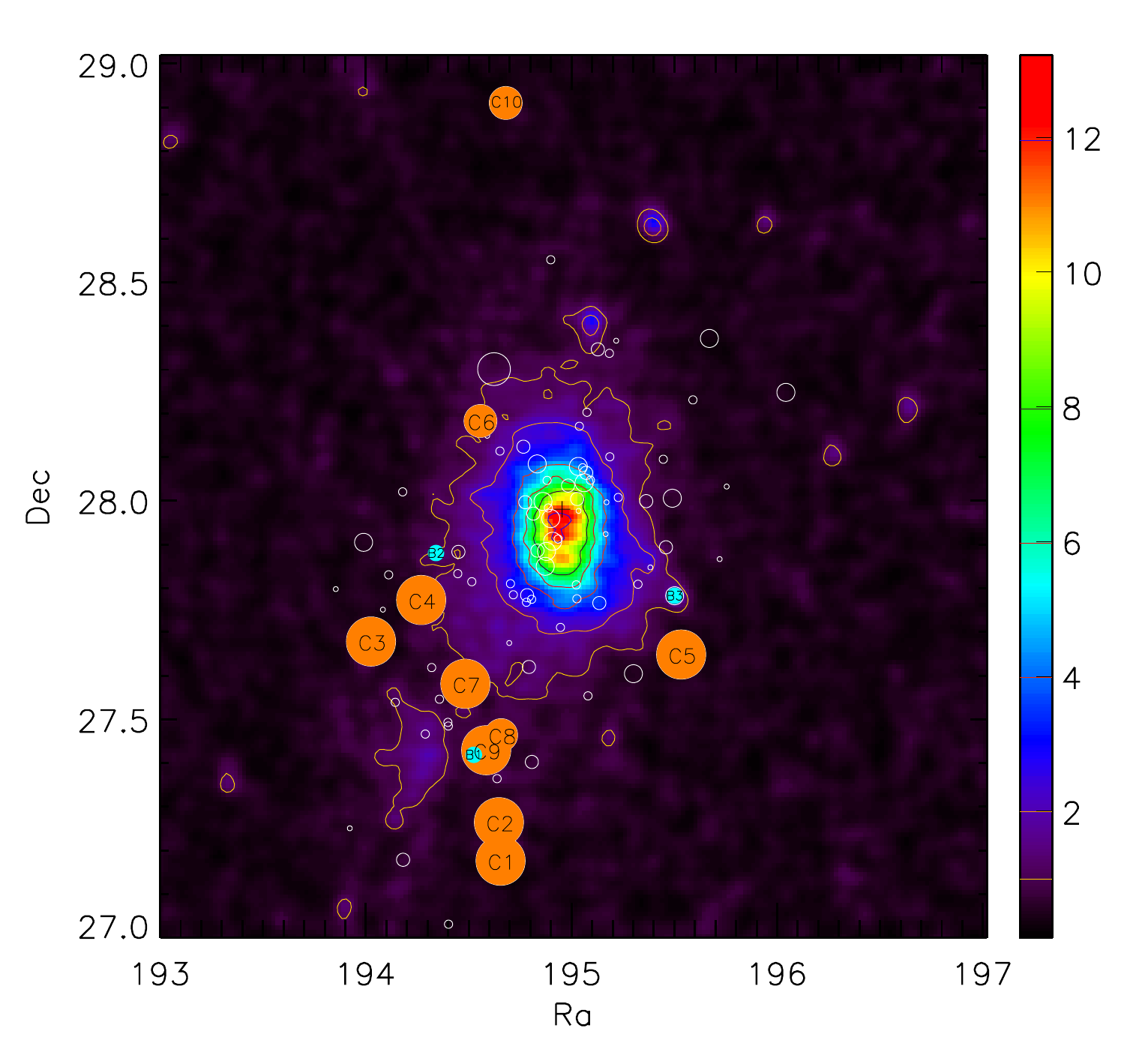}}
  \caption{Distribution of the 24$\mu$m detected sources (white open circles, orange and blue filled ones) on top of the ROSAT map of the Coma cluster. Orange filled circles represent the excess sources described in the text and populate mainly the southwest region of the cluster known to undergo a merger from an infalling substructure. The size of the circle is proportional to the ratio L$_{24\mu m}$/L$_K$ (i.e proportional to the specific star-formation rate of the source). The 6 blue sources are the ``blue ETGs'' that we define in sec.\,\ref{sec:opt} and describe in sec.\,\ref{sec:opt}\,\&\,\ref{sec:spec}. The color bar on the right is the X-ray temperature in keV. The color of the filled circles are not linked to this color bar neither to the temperature.}
 \label{fig:bubble_plot_coma}
\end{figure*}

Fig.\,\ref{fig:bubble_plot_coma} shows the distribution of the ETGs in the Coma cluster with respect to their 24$\mu$m/K-band luminosities ratio. Indeed, the circles are drawn proportionally to the L$_{24}$/L$_K$ ratio of the sources. MIEGs are marked in orange on the figure.
The ETGs are equally distributed between the core and the south region of the cluster ($\sim$40\% of the sources from our Coma sample lie in each region) and only 20\% of our sample lie in the north part of the cluster. In the core of the cluster, the whole ETGs population is non-MIEG i.e. does not show signs of star formation, while 20\% of the ETGs are MIEGs in the soutwest region (7 sources among the 10 MIEGs) against only 1\% of the ETG sample being selected as MIEG in the north part of the cluster. The fact that the MIEGs are mainly located in outskirt of the galaxies and do not populate the core is expected since it is known that the star-forming galaxies preferentially avoid the core of the clusters and populate the outskirts \citep[e.g.,][]{Kennicutt:83,Dressler:99,Gallazzi:09}. The exciting result here is that the MIEGs are not randomly distributed in the outskirts of the cluster but are mainly located in the southwest region
of the Coma cluster, region known to experience an ongoing merger and a lot of thermodynamic activity. \citet{Briel:92,Watt:92} argue, from X-ray observations, that a gravitationally bound substructure may be falling into the Coma cluster.
\citet{Neumann:01} showed, using X-ray data, that this southwestern substructure is in falling into the main cluster. \citet{Simionescu:13} show that this ongoing merger affects the surface radio profile of the cluster in this direction. 
\citet{Caldwell:93} found that galaxies with signs of star formation (strong Balmer-line absorption) are mainly located between the main part of the Coma cluster and the southwestern substructure, leading to the conclusion that the merger is triggering the star formation in these sources. Two scenarios have been proposed over the past decades and are still debated. One is in favor of infalling substructure triggering the star formation due to the increasing external pressure \citep[e.g.,][]{Dressler:83,Evrard:91,Abraham:96,Wang:97,Moss:00,Gavazzi:03,Poggianti:04}. The second scenario invokes gas stripping in galaxies due to ram-pressure in the ICM that will quench the star-formation in galaxies during mergers \citep[e.g.,][]{Tomita:96,Balogh:97,Balogh:98,Fujita:99,Baldi:01}. Our results on the MIEGs would be more in favor of the former effect and we argue that the infall of the substructure could enhance star-formation in some ETGs. 
\citet{Ferrari:05} found similar results in Abell 3921 from a {\it VRI} photometric and spectroscopic survey of the cluster. They argue that the ongoing merger in Abell 3921 has triggered a new star-formation episode in a non negligible fraction of their sample of emission-line galaxies. \\
In the Virgo cluster, we observe the same expected trend where MIEGs are distributed mainly in the outskirts of the cluster, beyond the virial radius. Two of the three Virgo MIEGs (V1\& V3) are located at the border between cluster B and the W' cloud that is thought to infall onto the cluster B. Given the small number of MIEGs in Virgo and the large number of significant substructures in Virgo, it is hard to argue whether Virgo MIEGs support the idea that MIEGs are located in merging area or not. Virgo MIEGs seem at least to not be spatially distributed like normal ETGs, since most ETGs are within or in the proximity of the cluster A area, whereas most Virgo MIEGs are in the outskirts of the cluster.

\subsection{Optical vs Infrared color color plot: comparison with late type}
\label{sec:opt}

\begin{figure}[h]
 \centering
  \resizebox{1.\hsize}{!}{\includegraphics{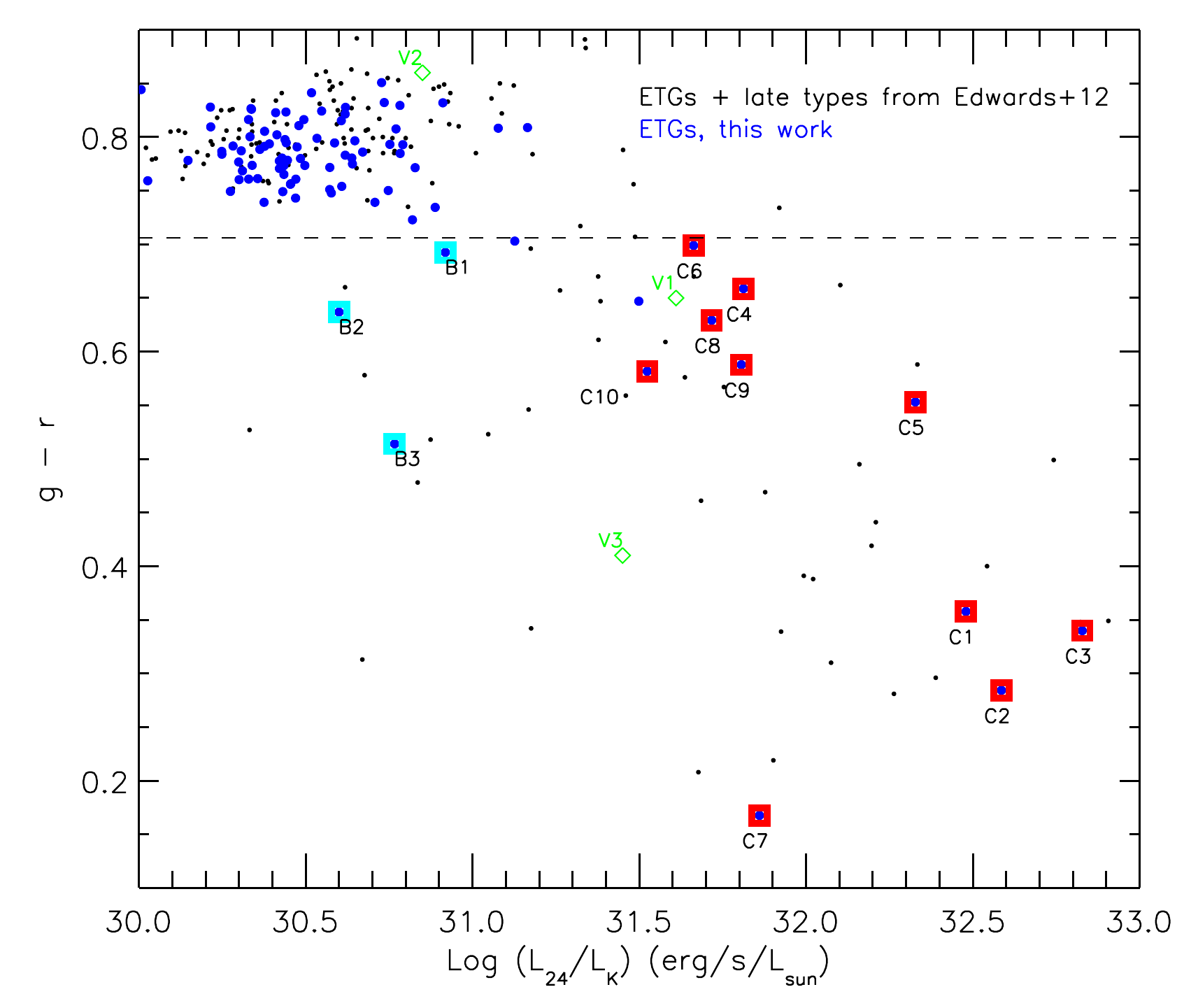}}
  \caption{optical g-r color as a function of the 24$\mu$m-to-K-band flux ratio for Coma sources:  our Coma ETG sample (blue dots) with the excess sources marked in green (Virgo) and red (Coma) and the ``blue ETGs''  in cyan and compared to the galaxies (both early and late-type, black dots) from \cite{Edwards:11}. }
 \label{fig:late_type_2}
\end{figure}

As discussed in the previous subsections, the extreme sources exhibit different properties than the rest of the ETG sample. 
The morphology of the MIEGs, described in sec.\,\ref{sec:morph}, confirm them as early-type sources. However their large L$_{24}$/L$_K$ ratio encourage us to compare MIEGs to late-type galaxies  to understand their evolutionary pathway. 
Fig.\,\ref{fig:late_type_2} displays the optical (g-r) color as a function of the 24$\mu$m-to-K-band flux ratio of our 24$\mu$m-detected ETG sample (blue dots) with the 13 MIEGs encircled in green and red and compared to late-type and early-types sources in Coma from \citet{Edwards:11}. 
The general conclusion from these figures is that MIEGs clearly show a different behavior than normal ETGs as they are located at a different position in the color-color diagram. In Fig.\,\ref{fig:late_type_2}, MIEGs have a lower (g-r) color compared to the rest of the ETG sample, underlying a potential excess in the g band. Four MIEGs (C1, C2, C3 \& C7) have far bluer colors than any other ETGs. Only late-type sources are as blue as these 4 MIEGs and with similar 24$\mu$m-to-K-band flux ratios. V2, one of the Virgo MIEGs, is located with the bulk of ETGs way above the other MIEGs in terms of  g-r colors, it is also the galaxy with the lowest \rattk\, ratio and was barely above our selection line.\\ 
Three ETGs (marked as B1 to B3 on Fig.\,\ref{fig:late_type_2} and encircled in cyan) with F$_{24}$/F$_K$\,$<$\,0.4 (i.e. not MIEGs) have bluer color than the bulk of the ETGs, similar to the g-r colors of some MIEGs. These three sources are all lenticulars.
The 3 ``blue ETGs'' seem to occupy areas of higher X-ray surface brightness and are located near a MIEG (looking at the distribution of the blue and orange sources on Fig.\,\ref{fig:bubble_plot_coma}). They do no show sign of enhanced star-formation but present a bluer color than an ordinary early-type galaxy.
\citet{Caldwell:93} have identified some ETGs in Coma with an abnormal spectrum mainly located in the SW field.
They found that their abnormal-spectrum galaxies are slightly bluer (in B-V) than their normal counterparts. We found the same result but looking at the g-r color which is similar to the B-V used by \citet{Caldwell:93}. 
We are providing evidence from the MIR photometry of likely the same phenomenon as \citet{Caldwell:93}. These ``blue'' ETGs are therefore reliable candidates for being post-MIEG where the star-formation would have been quenched by ram pressure stripping by the ICM. Our ``blue'' galaxies are similar to the E+A galaxies from the literature \citep[or more recently called ``k+a'' and ``a+k'' galaxies, e.g.,][]{Dressler:83,Franx:93} who have undergone a recent star formation activity followed by a quiescent phase. The southwest region of the Coma cluster is as an fine laboratory where we can observe transition-phase galaxies at different stages of their evolutionary path.

\subsection{Analysis of the SDSS spectra}
\label{sec:spec}

\begin{figure*}
% \centering
 \resizebox{1.\hsize}{!}{\includegraphics{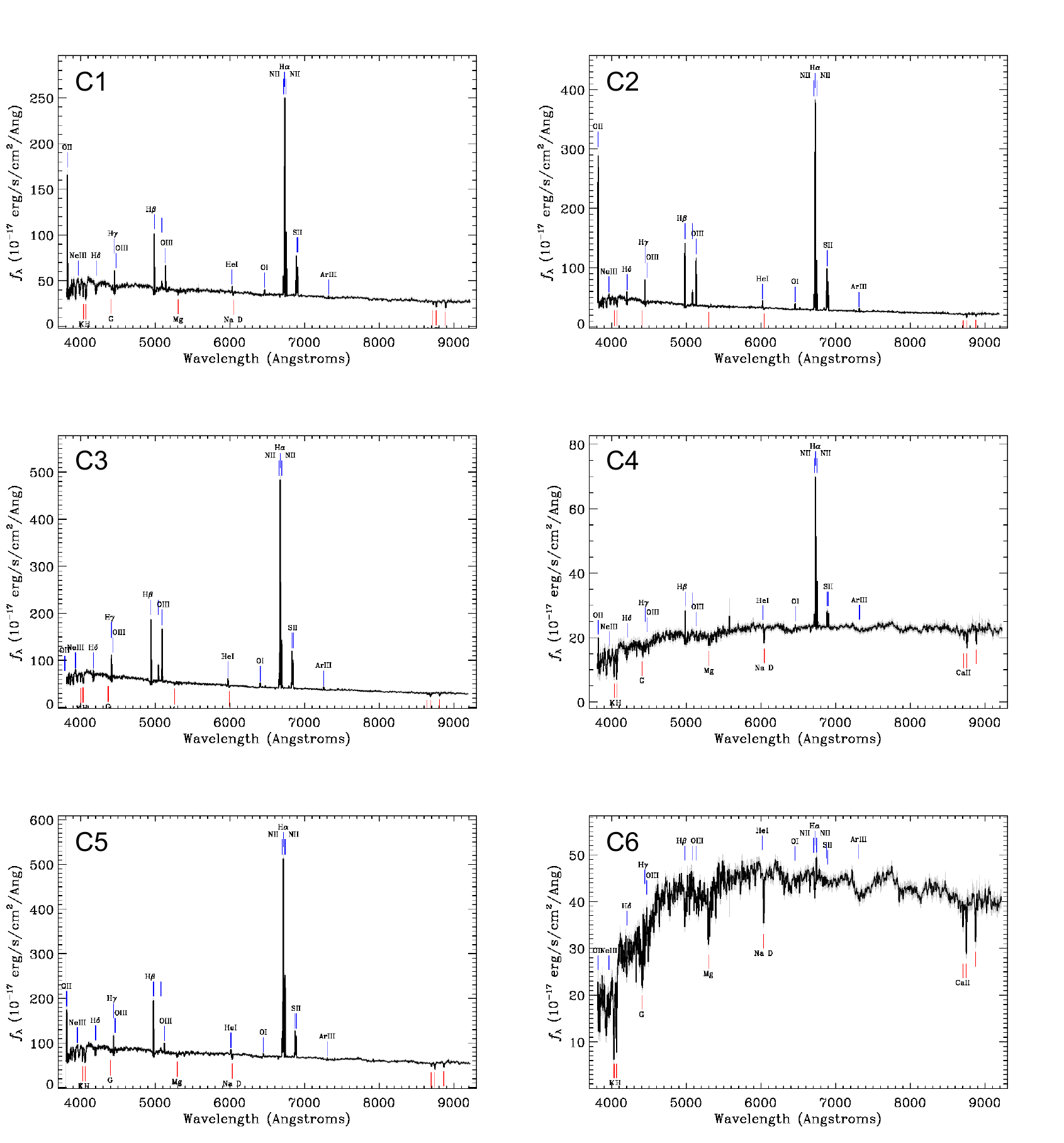}}
  \caption{SDSS spectra of the first six Coma MIEGs (C1-C6).}
 \label{fig:spectra1}
\end{figure*}

\begin{figure*}
  \resizebox{1.\hsize}{!}{\includegraphics{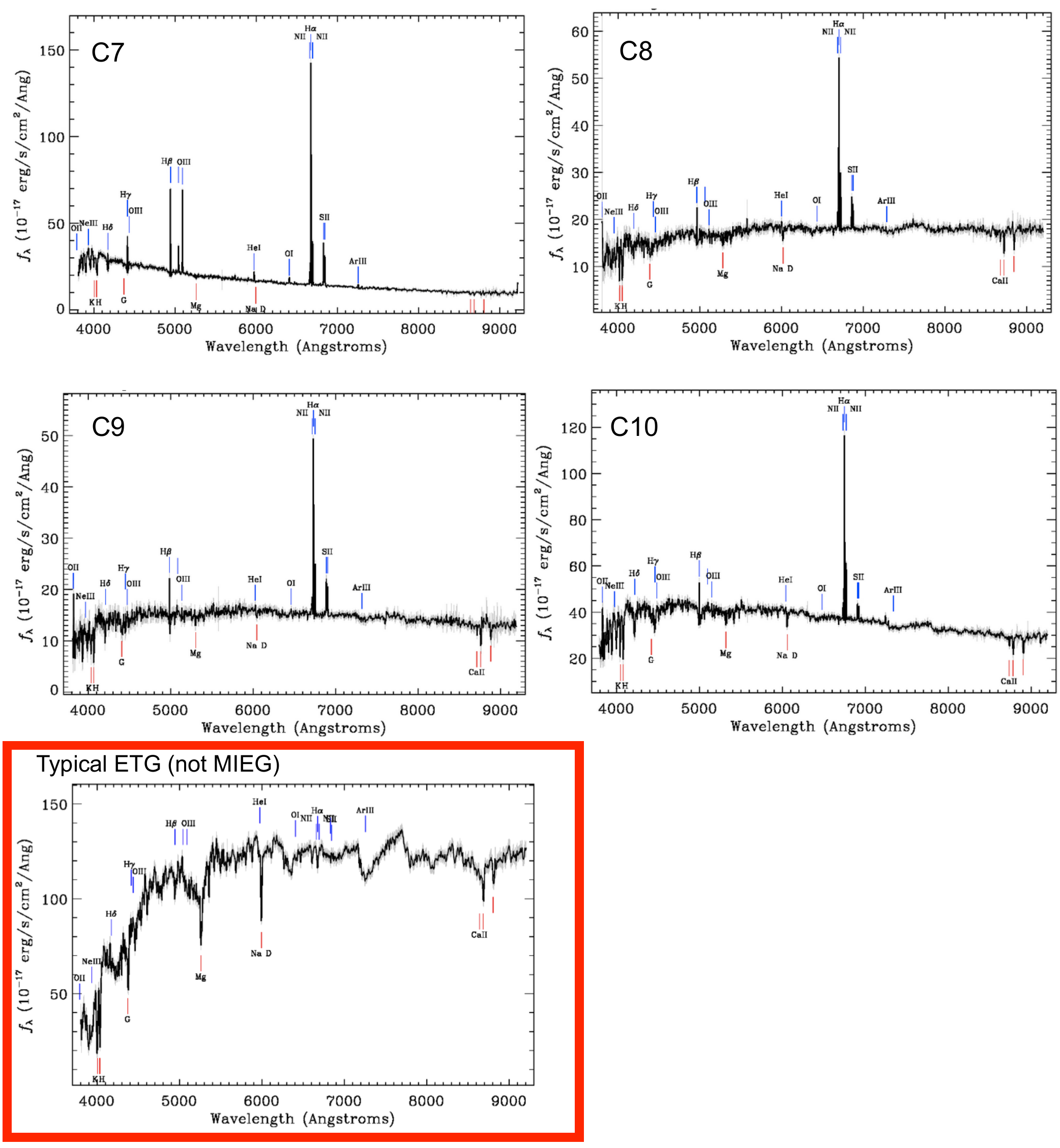}}
  \caption{SDSS spectra of the last four Coma MIEGs (C7-C10). The last spectrum on the bottom left is a typical spectrum of an ETG in Coma which is not a MIEG (the spectrum corresponds to the source framed in red in Fig.\,\ref{fig:SDSS}.}
 \label{fig:spectra2}
\end{figure*}

\begin{figure}
% \centering
 \resizebox{1.\hsize}{!}{\includegraphics{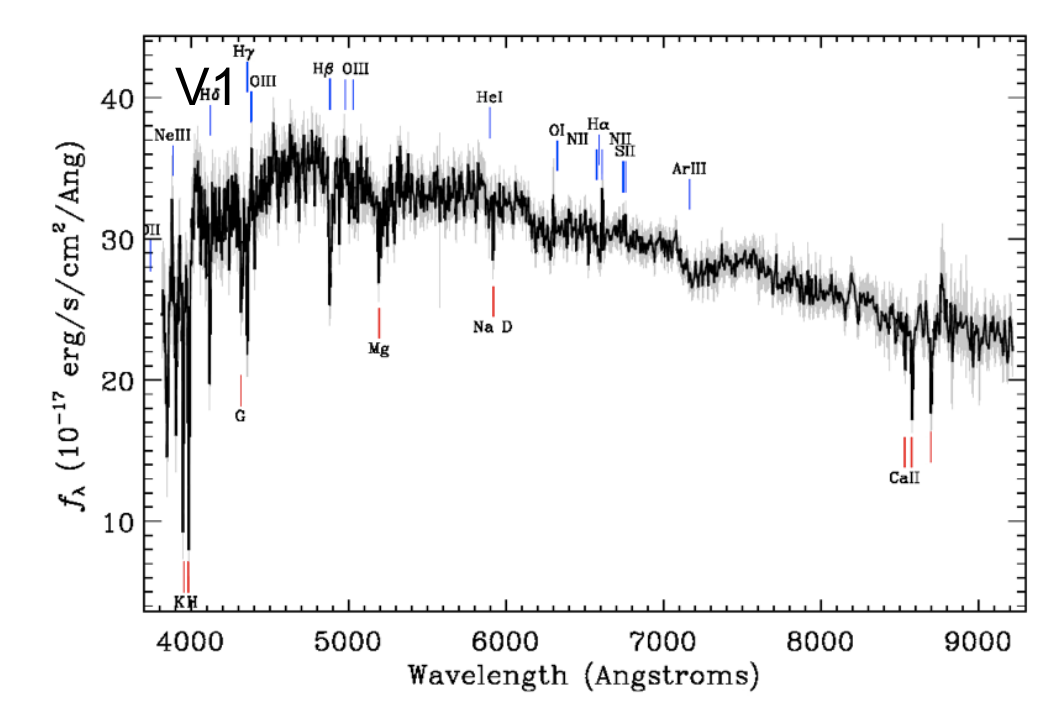}}
  \caption{SDSS spectra of 1 Virgo MIEGs (V1).}
 \label{fig:spectra_v}
\end{figure}

\begin{figure*}
% \centering
 \resizebox{1.\hsize}{!}{\includegraphics{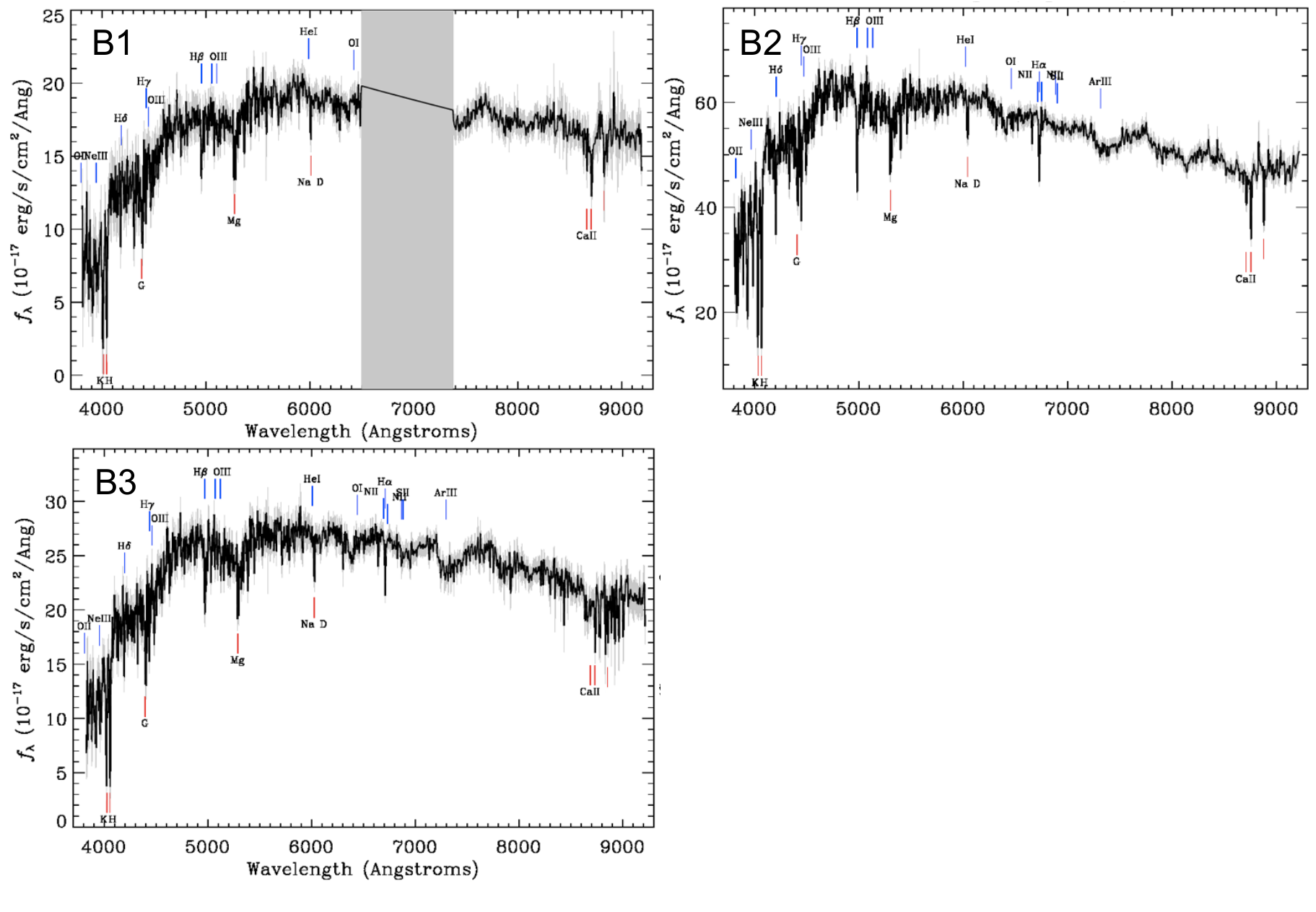}}
  \caption{SDSS spectra of the 3 ``blue ETGs'' (noted B1 to B3 on Fig.\,\ref{fig:late_type_2}).}
 \label{fig:spectraB}
\end{figure*}

We obtain the SDSS spectra for 12 MIEGs (see Fig.\,\ref{fig:spectra1}\,\&\,\ref{fig:spectra2},\&\,\ref{fig:spectra_v}). The 3'' fiber diameter (displayed as a red circle on Fig.\,\ref{fig:SDSS}\,\&\,\ref{fig:SDSS_v}) for the SDSS spectrograph (i.e. $\sim$1.4\,kpc at Coma's distance but only 0.24\,kpc at Virgo's distance) probes the central region of Coma MIEGs and a small inner fraction of that area for Virgo MIEGs. Given the smallness of the area probed, the interpetation of Virgo MIEG spectra has to be taken with caution, but is included here for completeness.\\
To compare with normal ETGs, we also obtained the SDSS spectra of a control sample, i.e. 10 ETGs in the Coma cluster that are not MIEGs. These spectra show globally the same features, i.e. no emission lines and a continuum increasing with wavelength. A typical ETG spectra is shown on Fig.\,\ref{fig:spectra2} on the bottom-right panel (encircled in red) as an indication. \\
C1, C2, C3, C5 \& C7 present a similar spectrum with a rather blue continuum. This is confirmed by their g-r color on Fig.\,\ref{fig:late_type_2} where C1, C2, C3 \& C7 exhibits bluer colors than any other ETGs from the sample (g-r $<$ 0.35). C5 also presents a rather blue g-r color (g-r $\sim$ 0.55) while the bulk of the ETG population (red points) lies at g-r $\sim$0.8. \\
C4, C8, C9, C10, V1 has a flatter  continuum and its g-r color is only slightly bluer than the bulk of the ETG. \\
C6 spectrum continuum is even more flat and resemble the one of a typical ETG.

All sources but C6 \& V1 exhibit a spectrum with strong emission lines. Indeed, all the sources, but these 2, have a strong H$_{\alpha}$ double emission lines, a SII emission line (around 6\,800\,\AA) and a H$_{\beta}$ emission line, while C6 \& V1 show a H$_{\beta}$ absorption lines. C1, C2, C3, C5 \& C7 have also a strong H$_{\gamma}$ emission line and 
H$_{\delta}$ is mostly seen as an absorption line except for C2. V1 lack of emission lines could be due to the 
smallness of SDSS beam, SDSS images show for instance that blue regions for V1 are located 2.2'' away from its center.

Three ETGs present g-r color as blue as some MIEGs (the ETGs encircled in cyan and marked as B1 to B3 on Fig.\,\ref{fig:late_type_2}). Fig.\,\ref{fig:spectraB} present the spectra of these 3 ``blue ETGs'', it is significantly different from the MIEGs and more similar to ``typical ETGs'' with a slightly bluer continuum (and also quite similar to C6 \& V1). This result supports the idea that the blue sources are on an advanced stage of the evolutionary path of the MIEGs and are therefore good post-MIEG candidate.

\begin{figure*}%[htpb]
 \centering
 \resizebox{0.5\hsize}{!}{\includegraphics{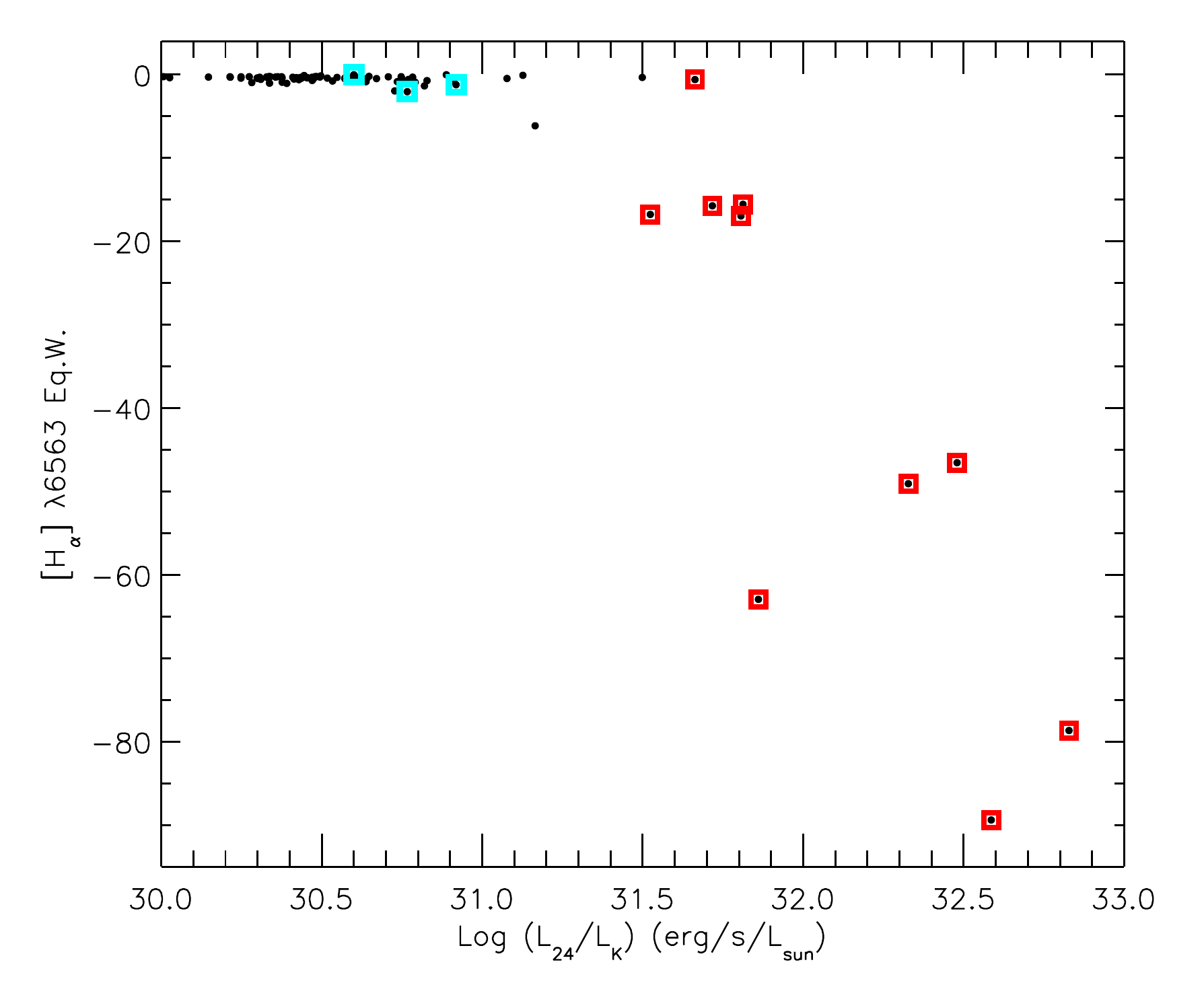}}
\resizebox{0.49\hsize}{!}{\includegraphics{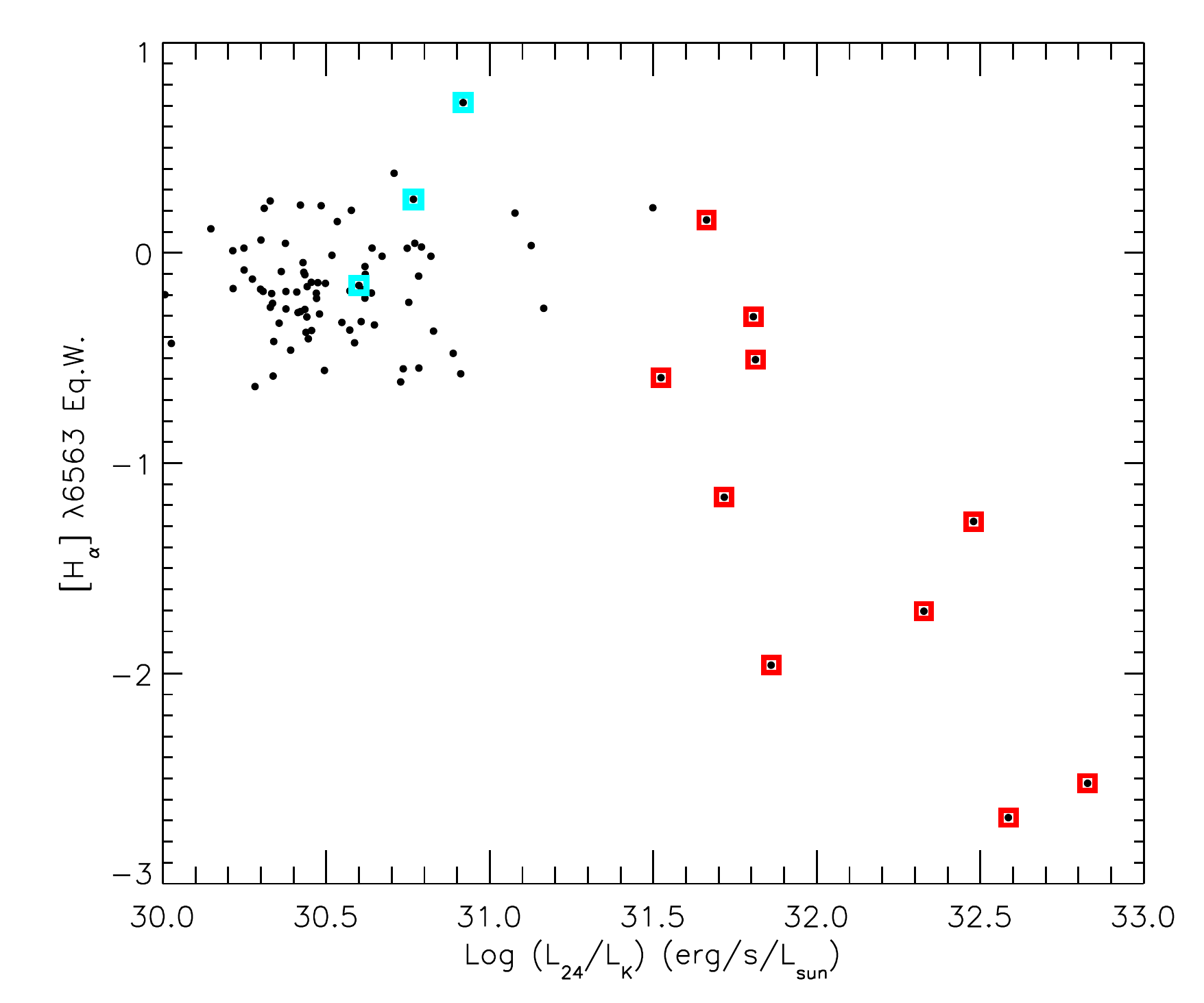}}
  \caption{{\it Left:} [H$_{\alpha}$]$\lambda$6563 versus F$_{24}$/F$_{K}$ for our Coma ETGs (emission is negative, absorption is positive). MIEGs are framed by red squares and ``blue ETGs'' are framed by blue squares.  {\it Right: } [H$_{\delta}$]$\lambda$4102 versus F$_{24}$/F$_{K}$ for our Coma ETGs with the same color coding as the left panel.}
 \label{fig:spec_lines}
\end{figure*}

In Fig.\,\ref{fig:spec_lines}, the equivalent widths of H$_{\alpha}$ and H$_{\delta}$ spectral emission lines are plotted as a function of the 24$\mu$m/K flux ratio.
The equivalent widths of the continuum-subtracted emission lines are obtained from the SDSS database and have been computed from straight integration over the corresponding bandpasses (emission is negative) . Stellar absorption is taken into account.

A correlation is observed for both H$_{\alpha}$ and H$_{\delta}$ spectral emission lines with respect to the F$_{24}$/F$_{K}$ ratio when the ratio is larger than about 0.8. At low F$_{24}$/F$_{K}$, no correlation with H$_{\alpha}$ and H$_{\delta}$  is observed for normal Coma ETGs (black dots), including the ``blue'' sources (black dots with a blue square). These two panels of Fig.\,\ref{fig:spec_lines} confirm the peculiarity of the MIEG spectra. The strong correlation at higher F$_{24}$/F$_{K}$ can be explained by the fact that H$_{\alpha}$ and H$_{\delta}$ are indicators of star-formation activity, and such a correlation between F$_{24}$/F$_{K}$ and the star-formation has already been observed by \citet{Temi:09b}.

\section{Discussion}
\label{sec:discussion}

Galaxies with a large L$_{24}$/L$_K$ ratio for ETGs, referred to as MIEGs, Mid-Infrared Enhanced Galaxies, are observed in both the Coma and Virgo clusters. In the Coma cluster, MIEGs are located primarily (70\% of them) in the merging area of the central structure of the cluster and its main substructure. In the Virgo cluster, the location of the MIEGs is more spread, but their spatial distribution
do not follow the one of typical ETGs. Virgo MIEGs are predominantly (80\% of them) located in the outskirts of the cluster, some in the vicinity of merging area, but it is not clear from their location if all MIEGs share the same origin. HST images, when available, show that MIEGs have disturbed
but not identical morphologies, more images would be necessary to assess if morphologies indicate a common origin.

\citet{Caldwell:93} already found an abnormal sample of early-type galaxies in Coma looking at 
their optical spectra obtained on the KPNO multifiber positioner, Hydra (3.8 angstrom resolution and
2'' fiber fov). Their sample of 125 ETGs concentrated on the central part of the cluster and the SW region,
with roughly 49 E and 76 S0 galaxies. 
\citet{Caldwell:93} found 3 abnormal galaxies in the central area out of 74 galaxies and 14 abnormal
 galaxies in the SW region out of 51 galaxies, galaxies were classified as abnormal when their CN/H8 flux slope
index is quite negative and their H$\delta$ equivalent width is large. 15 Galaxies among these 17 have 
enhanced Balmer absorption lines.
\cite{Caldwell:93} showed that these sources are not misclassified spiral galaxies and that their spectra 
are closer to the ones of E+A post-starburst galaxies found at intermediate redshifts, and even closer to 
red H$\delta$-strong galaxies found by \citet{Couch:87} in several clusters at z$\simeq$0.3.\\
57 galaxies in our sample are common to the \cite{Caldwell:93} sample, 8 galaxies among the abnormal ETGs are not 
in our sample, 3 galaxies in both sample are qualified as abnormal by \citet{Caldwell:93} and as MIEGs in this work, 
6 galaxies are classified as abnormal by
\citet{Caldwell:93} but not as MIEGs in our work. Among these 6 galaxies, 4 have been removed from our analysis
because they have been identified as AGN dominated with their optical spectra, the 2 others, one elliptical 
and one lenticular, did not have a very large  \rattk\, ratio (30.4 and 30.6) and are located in the bulk of 
the Coma ETG population on the L$_{24}$ vs L$_K$ diagram. Their SDSS g-r colors are mildly blue (0.54 and 0.64),
and we identified them with our non-MIEG blue ETGs B1 and B4. The galaxies identified by \citet{Caldwell:93}
as ETGs with an abnormal spectrum are therefore not all MIEGs, but it seems possible that MIEGs are a subcategory
of these ETGs with an abnormal spectrum. Knowing that the 2 sources identified by \citet{Caldwell:93} are blue but
do not show a large SFR could indicate that they are post-starburst and have exhausted their gas. Our unique mid-to-near IR analysis have allowed us to add 7 ``abnormal'' sources to the study of \citet{Caldwell:93}. In this picture,
MIEGs would be starburst ETGs and \citet{Caldwell:93} abnormal ETGs would be composed of these starburst ETGs and
also of post-starburst ETGs.  \\

\citet{Davis:14} investigated the position of ETGs on the Kennicutt-Schmidt relation, using a star-formation 
rate estimated with WISE 22 $\mu m$ and also found some sources with a strong \rattk\, ratio. In order to remove 
the contribution from the emission of circumstellar dust, they looked at the same correlation as \citet{Temi:09b} 
but with WISE data, i.e. L$_{22\mu m}$ vs L$_{K_s}$, for all 260 galaxies of the ATLAS$^{3D}$ sample. Using molecular 
gas poor galaxies (no CO detection), they measured the L$_{22\mu m}$--L$_{K_s}$ relation due to the circumstellar 
dust. 
While most molecular gas poor ETGs lie within the photometric uncertainties of the fitted relation, ETGs with a 
CO detection have an excess mid-infrared luminosity and do not show any correlation between L$_{22\mu m}$ 
and the stellar luminosity (i.e., L$_{K_s}$ since the K$_s$ band is a proxy for stellar mass). 
From their derivation of the SFR, they emphasized that molecular-rich ETGs present higher SFR densities 
than disks of spiral galaxies but are comparable to the densities found at the center of spiral galaxies.
They argued that the dynamical stability of the gas could play a role in the suppression of star-formation in ETGs. 
These molecular-rich galaxies from the ATLAS$^{3D}$ survey have a 
L$_{22\mu m}$/L$_{K_s}$ ratio similar to our MIEGs, but it is not clear whether they lie in a cluster-substructure merging
area like our MIEGs. \citet{Serra:12} studied the correlation of the HI ETG mass and the environment of the 
ATLAS$^{3D}$ galaxies and found that gas-rich galaxies live in the poorest environments whereas galaxies in the
richest environment (center of the Virgo cluster) have the lowest HI content. MIEGs, which lie at the border
between the central cluster and an infalling substructure, sit in an intermediate environment. On average,
\citet{Serra:12} found that the HI mass of galaxies in an intermediate environment is in between the HI mass in 
the poor and rich environment, but the HI mass versus environment density relation is quite spread and the 
environment density is not a precise indication of the amount of substructure merger activity.  CO-rich galaxies 
in \citet{Davis:14} and MIEGs have a large \rattk\, common ratio but it is not clear whether they share the same origin.

Using {\em Suzaku} observations, \citet{Simionescu:13} measured the temperature radial profile of the 
intracluster medium (ICM) hot gas along the Southwestern direction and in other directions not
 aligned with the infalling subcluster. 
In the SW direction, they found a steady decline of the ICM temperature outside of the cluster core ($\sim$35-55 arcmin), 
from 8.5 keV to 4 keV,  a constant temperature around 55-78 arcmin and a dramatical drop in the last annulus (78-88 arcmin), which may be the sign of the boundary of a shock. 
MIEGs lie in the 70-100 arcmin range, close to the Coma 2.84 Mpc virial radius, since 1 arcmin 
at the distance of Coma is equivalent to 28\,kpc. 
At this distance, the temperature in the SW direction is lower than around the cluster core
but it is higher than in any other direction (where there is no infalling substructure). 
Moreover, from the cluster center to $\sim$65\,arcmin, the temperature profiles along the SW direction and 
along other directions are globally the same.
Using {\em Chandra} observation of Abell 521 (z$\simeq$ 0.247), a rich cluster undergoing several mergers,
\citet{Ferrari:06} observed a bar-shaped high ICM temperature (5 to 7 keV) located between 
the northern and southern regions of the cluster. They argued that the gas could have been 
compressionnally heated by the infalling northern substructure, in agreement with \citet{Simionescu:13}
observation of Coma.

\section{Summary and Conclusions}

This paper analyzed a sample of 25 ETGs in the Virgo cluster and 95 ETGs in the Coma cluster down to the same data depth (F$_{24\mu m} >$ 12 mJy, K$_s <$ 10.53 [Vega] and  r $<$ 12.9 [AB] ) to compare the mid-IR properties of ETGs in two rich environments at a different stage of their evolution. We focus on the L$_{24\mu m}$ /L$_{K}$ distribution of the sources, a proxy for the star formation activity and we also obtain the optical colors, the detailed morphology and the optical spectra for a subset of our sample.
Using these sets of data, we derived the following conclusions:

\begin{itemize}
\item The L$_{24\mu m}$ /L$_{K}$ distribution of ETGs in Coma peaks at a higher value than for Virgo ($\sim$0.30\,$\pm$\,0.08\,dex difference). 
\item 13 ETGs with signs of strong star-formation activity (with a 24 $\mu$m flux well above the expected value) are all lenticulars. We named them Mid-Infrared Enhanced Galaxies (MIEGs). MIEGs exhibit diverse morphologies, from globally, asymmetric systems to redder, more relaxed galaxies. The majority of the MIEGs in Coma (7 sources among the 10 ones) are located in the same neighborhood of the cluster, the southwest part of Coma, known for an ongoing merger activity. MIEGs also have bluer g-r color than the rest of the ETG population, with g-r colors similar to late-type galaxies. Their spectra are different from typical ETG spectrum, and feature strong emission lines (e.g., H$_{\alpha}$, H$_{\delta}$, OII) that are good indicators of star-formation. A tight correlation is also observed for the MIEGs between the H$_{\alpha}$ and H$_{\delta}$ emission lines equivalent widths and the F$_{24\mu m}$ /F$_{K}$ ratio, while ``normal'' ETGs do not show a correlation between their emission lines equivalent widths and their F$_{24\mu m}$ /F$_{K}$ ratio.
\item Three ETGs in Coma present similar g-r colors than some of the MIEGs, they are bluer than the bulk of the ETG population because of their blue continuum. They do not show any strong emission lines that could account for star-formation activity like MIEGs. However these ``blue sources'' are all located in the neighborhood of MIEGs that hint toward a common history and a common evolution different from the rest of the ETG population.
\end{itemize}

These results emphasize the potential link between cluster merger activity and star-formation in galaxies. From a mid-to-near IR analysis, we have selected transiting galaxies at different stages of their evolutionary history. We have selected ETGs with signs of strong star formation and a subclass of ETGs with blue optical colors but no remaining of current star formation. 
MIEGs are a peculiar population of lenticular galaxies with signs of enhanced star formation (high L$_{24\mu m}$ /L$_{K}$ ratios, strong emission lines, blue optical colors) potentially located in cluster substructure merging area. The known ongoing merger activity of the Coma region, where most of Coma MIEGs are located, could account for this enhancement. The hypothesis of cluster merging triggering the star formation has been widely discussed and accepted \citep[e.g.,][]{Bekki:99,Gavazzi:03,Poggianti:04}, mostly for starbursts. In the other hand, the blue sources discussed in this work are good post-MIEG candidates where the recent star formation would have been quenched due to ram pressure stripping \citep[e.g.][]{Baldi:01}. These blue sources would be the analogs of the post-starburst candidates in a evolutionary path for ETGs where MIEGs would be the analog of starburst galaxies.
We will focus in the near future on the formation of these MIEGs, analyzing for instance their stellar population and looking for a second burst of star formation.

\acknowledgements

This work is based on archival data obtained with the Spitzer Space Telescope, which is operated by the Jet Propulsion Laboratory, California Institute of Technology under a contract with NASA. Support for this work was provided by NASA.
This publication makes use of data products from the Two Micron All Sky Survey, which is a joint project of the University of Massachusetts and the Infrared Processing and Analysis Center/California Institute of Technology, funded by the National Aeronautics and Space Administration and the National Science Foundation. This publication makes use of data from SDSS-III. Funding for SDSS-III has been provided by the Alfred P. Sloan Foundation, the Participating Institutions, the National Science Foundation, and the U.S. Department of Energy Office of Science. The SDSS-III web site is http://www.sdss3.org/. SDSS-III is managed by the Astrophysical Research Consortium for the Participating Institutions of the SDSS-III Collaboration including the University of Arizona, the Brazilian Participation Group, Brookhaven National Laboratory, University of Cambridge, Carnegie Mellon University, University of Florida, the French Participation Group, the German Participation Group, Harvard University, the Instituto de Astrofisica de Canarias, the Michigan State/Notre Dame/JINA Participation Group, Johns Hopkins University, Lawrence Berkeley National Laboratory, Max Planck Institute for Astrophysics, Max Planck Institute for Extraterrestrial Physics, New Mexico State University, New York University, Ohio State University, Pennsylvania State University, University of Portsmouth, Princeton University, the Spanish Participation Group, University of Tokyo, University of Utah, Vanderbilt University, University of Virginia, University of Washington, and Yale University.
\appendix
{\LongTables
\begin{deluxetable*}{lcccccccccccc}

\tablecaption{Early Type Sample for the Virgo Cluster}
\tablewidth{0pt}

\tablehead{
Name  &  VCC & Dist  & T  &  $F_{24\mu m}$ &  $ \Delta F_{24\mu m}$  & $Ks$ &  $\Delta Ks$  &  AGN  & 
log $L_{24\mu m}$ &  log $L_{Ks}$ &  $ W4$ &  $\Delta W4$\\
\colhead{}  & \colhead{}  &  \colhead{(Mpc)}  &  \colhead{}   &  \colhead{(mJy)}  &  \colhead{(mJy)}  &  \colhead{(Vega mag)}  &  
\colhead{}  &  \colhead{}  & \colhead{(erg s$^{-1}$)}  & \colhead{(L$_{\odot}$)}  &  \colhead{(Vega mag)}  & \colhead{}\\
\colhead{(1)}  &  \colhead{(2)}   &  \colhead{(3)}   &  \colhead{(4)}  &  \colhead{(5)}  &  \colhead{(6)}   &  \colhead{(7)}   &  
\colhead{(8)}  &  \colhead{(9)}   &  \colhead{(10)}  &  \colhead{(11)} &  \colhead{(12)} &  \colhead{(13)}
 }
 
\startdata
IC0798 & VCC1440 & 7.70 & -5.0 & 0.60 & 0.20 & 11.64 & 0.0750 &        0 & 38.72 & 8.46 & 10.29 & 2.01 \\
IC3328 & VCC0856 & 15.70 & -5.0 & 0.45 & 0.080 & 11.30 & 0.123 &        0 & 39.22 & 9.22 & 9.59 & 1.16 \\
IC3381 & VCC1087 & 10.90 & -4.1 & 0.52 & 0.100 & 11.05 & 0.119 &        0 & 38.96 & 9.00 & \ldots & \ldots \\
IC3468 & VCC1422 & 19.60 & -4.9 & 0.16 & 0.080 & 10.51 & 0.113 &        0 & 38.96 & 9.73 & \ldots & \ldots \\
IC3501 & VCC1528 & 24.80 & -3.5 & 0.24 & 0.080 & 11.24 & 0.0570 &        0 & 39.34 & 9.64 & 10.89 & 2.36 \\
IC3602 & VCC1743 & 31.40 & -5.0 & 0.69 & 0.11 & 11.84 & 0.0780 &        0 & 40.01 & 9.61 & 7.88 & 0.22 \\
IC3652 & VCC1861 & 10.30 & -4.9 & 0.22 & 0.080 & 11.05 & 0.127 &        0 & 38.54 & 8.96 & 10.13 & 1.72 \\
IC3653 & VCC1871 & 14.70 & -5.0 & 0.77 & 0.12 & 10.58 & 0.0500 &        1 & 39.40 & 9.45 & 9.44 & 0.97 \\
IC3773 & VCC2048 & 16.70 & -4.7 & 0.34 & 0.080 & 10.93 & 0.0700 &        0 & 39.15 & 9.42 & 9.96 & 1.78 \\
NGC4168 & VCC0049 & 33.70 & -4.8 & 5.20 & 1.30 & 8.44 & 0.0210 &        2 & 40.95 & 11.03 & 6.97 & 0.38 \\
NGC4261 & VCC0345 & 31.60 & -4.8 & 51.50 & 3.20 & 7.26 & 0.0280 &        2 & 41.89 & 11.44 & \ldots & \ldots \\
NGC4267 & VCC0369 & 15.90 & -2.7 & 11.70 & 0.30 & 7.84 & 0.0230 &        0 & 40.65 & 10.62 & 6.62 & 0.37 \\
NGC4344 & VCC0655 & 17.90 & -2.1 & 39.20 & 4.50 & 10.48 & 0.0240 &        1 & 41.27 & 9.66 & \ldots & \ldots \\
NGC4350 & VCC0685 & 15.90 & -1.8 & 29.80 & 3.30 & 7.82 & 0.00800 &        0 & 41.05 & 10.63 & 5.92 & 0.14 \\
NGC4352 & VCC0698 & 31.00 & -2.0 & 1.70 & 0.30 & 9.87 & 0.0450 &        1 & 40.39 & 10.38 & 8.01 & 0.57 \\
NGC4365 & VCC0731 & 17.10 & -4.8 & 22.20 & 4.70 & 6.64 & 0.0290 &        0 & 40.98 & 11.16 & 5.33 & 0.20 \\
NGC4371 & VCC0759 & 14.30 & -1.3 & 13.60 & 0.70 & 7.72 & 0.0230 &        0 & 40.62 & 10.57 & 6.10 & 0.23 \\
NGC4374 & VCC0763 & 18.40 & -4.3 & 66.60 & 8.60 & 6.22 & 0.0230 &        2 & 41.53 & 11.39 & 4.66 & 0.14 \\
NGC4377 & VCC0778 & 21.30 & -2.6 & 3.20 & 0.30 & 8.83 & 0.0160 &        0 & 40.33 & 10.47 & 5.93 & 0.12 \\
NGC4379 & VCC0784 & 14.20 & -2.8 & 4.30 & 0.30 & 8.77 & 0.0180 &        0 & 40.11 & 10.15 & 7.84 & 0.53 \\
NGC4382 & VCC0798 & 18.50 & -1.3 & 52.40 & 6.70 & 6.14 & 0.0210 &        0 & 41.43 & 11.42 & \ldots & \ldots \\
NGC4406 & VCC0881 & 17.10 & -4.8 & 27.50 & 3.20 & 6.10 & 0.0280 &        0 & 41.08 & 11.37 & 4.99 & 0.22 \\
NGC4417 & VCC0944 & 15.90 & -1.9 & 8.80 & 1.20 & 8.17 & 0.0280 &        0 & 40.52 & 10.48 & 6.40 & 0.21 \\
NGC4421 & VCC0966 & 23.10 & -0.3 & 3.30 & 0.40 & 8.80 & 0.0210 &        0 & 40.42 & 10.56 & \ldots & \ldots \\
NGC4434 & VCC1025 & 26.80 & -4.8 & 4.80 & 0.90 & 9.21 & 0.0290 &        0 & 40.71 & 10.52 & 8.06 & 0.47 \\
NGC4435 & VCC1030 & 15.90 & -2.0 & 111.00 & 10.20 & 7.30 & 0.0160 &        0 & 41.62 & 10.83 & 4.68 & 0.12 \\
NGC4442 & VCC1062 & 8.70 & -1.9 & 20.20 & 2.20 & 7.29 & 0.0240 &        0 & 40.36 & 10.31 & 5.62 & 0.13 \\
NGC4458 & VCC1146 & 17.20 & -4.8 & 3.30 & 1.10 & 9.31 & 0.0220 &        0 & 40.16 & 10.09 & 8.52 & 0.96 \\
NGC4459 & VCC1154 & 16.10 & -1.4 & 107.00 & 8.00 & 7.15 & 0.0110 &        0 & 41.62 & 10.90 & 4.44 & 0.072 \\
NGC4464 & VCC1178 & 15.90 & -0.9 & 2.00 & 0.30 & 9.58 & 0.0280 &        0 & 39.88 & 9.92 & 9.16 & 0.78 \\
NGC4472 & VCC1226 & 17.10 & -4.8 & 74.70 & 8.60 & 5.51 & 0.0170 &        0 & 41.51 & 11.61 & 4.10 & 0.15 \\
NGC4473 & VCC1231 & 15.70 & -4.7 & 26.30 & 6.70 & 7.16 & 0.0230 &        0 & 40.99 & 10.88 & 5.70 & 0.19 \\
NGC4474 & VCC1242 & 15.90 & -1.9 & 5.40 & 0.80 & 8.70 & 0.0150 &        0 & 40.31 & 10.27 & 7.12 & 0.30 \\
NGC4476 & VCC1250 & 17.20 & -3.0 & 35.70 & 4.10 & 9.47 & 0.0200 &        1 & 41.20 & 10.03 & 6.03 & 0.064 \\
NGC4477 & VCC1253 & 17.10 & -1.9 & 39.00 & 0.80 & 7.35 & 0.0130 &        0 & 41.23 & 10.87 & 5.53 & 0.17 \\
NGC4478 & VCC1279 & 18.10 & -4.8 & 12.60 & 3.90 & 8.35 & 0.0130 &        0 & 40.79 & 10.52 & 6.72 & 0.27 \\
NGC4479 & VCC1283 & 15.90 & -1.9 & 1.20 & 0.30 & 9.77 & 0.0300 &        1 & 39.66 & 9.84 & 8.59 & 0.79 \\
NGC4482 & VCC1261 & 27.80 & -4.8 & 0.30 & 0.16 & 10.58 & 0.0810 &        1 & 39.54 & 10.00 & 12.11 & 10.00 \\
NGC4483 & VCC1303 & 13.70 & -1.3 & 3.00 & 0.30 & 9.29 & 0.0320 &        1 & 39.92 & 9.91 & 8.05 & 0.64 \\
NGC4486 & VCC1316 & 16.10 & -4.3 & 154.00 & 9.00 & 5.81 & 0.0190 &        2 & 41.77 & 11.44 & 3.78 & 0.069 \\
NGC4489 & VCC1321 & 17.90 & -4.8 & 3.20 & 0.30 & 9.36 & 0.0280 &        1 & 40.18 & 10.11 & 8.27 & 0.54 \\
NGC4515 & VCC1475 & 15.90 & -3.1 & 1.90 & 0.20 & 9.89 & 0.0230 &        1 & 39.86 & 9.80 & 8.39 & 0.38 \\
NGC4526 & VCC1535 & 16.90 & -1.9 & 267.00 & 12.00 & 6.47 & 0.0200 &        0 & 42.06 & 11.21 & \ldots & \ldots \\
NGC4528 & VCC1537 & 20.80 & -2.0 & 4.40 & 0.70 & 8.97 & 0.0220 &        0 & 40.45 & 10.40 & 7.58 & 0.56 \\
NGC4552 & VCC1632 & 15.30 & -4.6 & 58.50 & 7.80 & 6.73 & 0.0240 &        0 & 41.31 & 11.03 & 5.12 & 0.14 \\
NGC4564 & VCC1664 & 15.00 & -4.8 & 23.90 & 3.70 & 7.94 & 0.0210 &        0 & 40.90 & 10.53 & 6.58 & 0.38 \\
NGC4570 & VCC1692 & 25.90 & -1.9 & 18.70 & 6.00 & 7.69 & 0.0130 &        0 & 41.27 & 11.10 & 6.04 & 0.35 \\
NGC4578 & VCC1720 & 18.50 & -2.0 & 5.20 & 0.90 & 8.40 & 0.0330 &        0 & 40.43 & 10.52 & 6.55 & 0.40 \\
NGC4612 & VCC1883 & 26.60 & -2.0 & 6.00 & 0.80 & 8.56 & 0.0260 &        0 & 40.80 & 10.77 & 6.93 & 0.32 \\
NGC4621 & VCC1903 & 18.30 & -4.8 & 34.90 & 6.30 & 6.75 & 0.0260 &        1 & 41.24 & 11.17 & 5.45 & 0.23 \\
NGC4623 & VCC1913 & 26.60 & -1.4 & 1.80 & 0.20 & 9.47 & 0.0370 &        0 & 40.28 & 10.41 & 6.93 & 0.32 \\
NGC4636 & VCC1939 & 17.10 & -4.8 & 31.80 & 5.60 & 6.42 & 0.0350 &        0 & 41.14 & 11.24 & 4.83 & 0.23 \\
NGC4638 & VCC1938 & 21.70 & -2.7 & 12.70 & 0.100 & 8.21 & 0.0190 &        0 & 40.95 & 10.74 & 7.04 & 0.68 \\
NGC4649 & VCC1978 & 17.10 & -4.6 & 108.00 & 10.00 & 5.74 & 0.0210 &        0 & 41.67 & 11.52 & 3.14 & 0.047 \\
NGC4660 & VCC2000 & 12.80 & -4.7 & 15.50 & 4.30 & 8.21 & 0.0170 &        1 & 40.58 & 10.28 & 6.43 & 0.17 \\
NGC4694 & VCC2066 & 18.20 & -2.0 & 110.00 & 9.00 & 8.95 & 0.0290 &        0 & 41.73 & 10.28 & 4.64 & 0.039 \\
NGC4754 & VCC2092 & 16.80 & -2.4 & 17.10 & 2.90 & 7.41 & 0.0300 &        0 & 40.86 & 10.83 & 5.82 & 0.22 \\
NGC4762 & VCC2095 & 15.90 & -1.8 & 39.10 & 6.50 & 7.30 & 0.0290 &        0 & 41.17 & 10.83 & \ldots & \ldots \\
NGC4262 & VCC0355 & 15.40 & -2.7 & 12.40 & \ldots & 8.36 & 0.0160 &        0 & 40.64 & 10.38 & \ldots & \ldots \\
VCC0571 & VCC0571 & 23.80 & -1.3 & 3.57 & \ldots & 12.48 & 0.160 &        0 & 40.48 & 9.11 & \ldots & \ldots \\
NGC4318 & VCC0575 & 22.10 & -5.0 & 5.87 & \ldots & 10.35 & 0.0400 &        0 & 40.63 & 9.90 & \ldots & \ldots \\
NGC4387 & VCC0828 & 18.00 & -4.9 & 5.64 & \ldots & 9.15 & 0.0200 &        0 & 40.43 & 10.19 & \ldots & \ldots \\
NGC4486B & VCC1297 & 16.30 & -5.0 & 2.16 & \ldots & 10.09 & 0.0170 &        1 & 39.93 & 9.74 & \ldots & \ldots \\
NGC4486A & VCC1327 & 18.30 & -5.0 & 6.06 & \ldots & 9.01 & 0.0150 &        0 & 40.48 & 10.27 & \ldots & \ldots \\
NGC4550 & VCC1619 & 15.50 & -2.1 & 9.80 & \ldots & 8.69 & 0.0190 &        0 & 40.55 & 10.25 & \ldots & \ldots \\
NGC4551 & VCC1630 & 16.10 & -4.9 & 5.91 & \ldots & 8.87 & 0.0180 &        0 & 40.36 & 10.22 & \ldots & \ldots 
\enddata
\tablecomments{Col. (1): Name.  
Col. (2): Entry in the Virgo Cluster Catalog \citep{Binggeli:85}.
Col. (3) Adopted distance. When available, values are taken from \citet{Tonry:01}, otherwise they are from the NASA Extragalactic 
Database (NED) and corrected for $H_0 = 70$ km s$^{-1}$ Mpc$^{-1}$.
Col. (4): Morphological T parameter obtained from the Hyperleda database \citep{Paturel:03}.
Col. (5) \& (6): {\em Spitzer} MIPS 24$\mu$m flux density and associated uncertainty from \citet{Temi:09b} for the 58 first sources, and from 
\citet{Leipski:12} for the last 8 sources.
Col. (7) \& (8): K$_s$ magnitude and associated uncertainty, obtained from 2MASS.
Col. (9): AGN classification flag, based partially on an optical BPT analysis. Unclassified source = 0, star forming galaxies = 1,
AGN = 2.
Col. (10): Log of the 24$\mu$m luminosity. Here $L_{\lambda}$ represents $\lambda L_{\lambda}$ in ergs s$^{-1}$, as in \citet{Temi:09b}. 
Col. (11): Log of the Ks-band luminosity in solar units. 
Col. (12) \& (13): WISE 22$\mu$m magnitude and associated uncertainty.
}
\end{deluxetable*}
}

{\LongTables
\begin{deluxetable*}{cccccccccccc}
\tablecaption{Early Type Sample for the Coma Cluster}
\tablewidth{0pt}

\tablehead{
RA  &  Dec  &  Dist  &  T  & $F_{24\mu m}$  & $\Delta F_{24\mu m}$  & $Ks$  &  AGN  & log $L_{24\mu m}$ &  log $L_{Ks}$ 
&  $W4$ & $\Delta W4$\\
\colhead{J2000}       & \colhead{J2000}  &  \colhead{Mpc}  &  \colhead{}   &  \colhead{(mJy)}  &  \colhead{}  &  
\colhead{(Vega mag)}  & \colhead{}       &  \colhead{erg s$^{-1}$}  &  \colhead{L$_{\odot}$}  &
\colhead{(Vega mag)}  & \colhead{}  \\
\colhead{(1)}  &  \colhead{(2)}  &  \colhead{(3)}   &  \colhead{(4)}   & \colhead{(5)}   &  \colhead{(6)}   &  
\colhead{(7)}  &  \colhead{(8)}  &  \colhead{(9)}   &  \colhead{(10)}  &  \colhead{(11)} &  \colhead{(12)}
 }
 
\startdata
 194.475  &  28.500  &  106.6  &  -4.4  &   0.043  &  0.088  &  12.41  &  0  &  39.86  & 10.44  &  \ldots  &  \ldots \\
 195.217  &  28.366  &  111.7  &  -3.1  &   1.389  &  0.099  &  10.82  &  0  &  41.41  &  11.11  &    9.38  &   0.88 \\
 194.899  &  28.551  &  110.9  &  -2.5  &   0.444  &  0.096  &  12.07  &  0  &  40.91  &  10.61  &   9.98  &   0.90 \\
 195.670  &  28.371  &  107.8  &  -0.2  &   2.885  &  0.143  &  11.33  &  0  &  41.70  &  10.88  &   9.04  &   0.63 \\
 195.536  &  28.387  &  110.8  &  -3.5  &   7.705  &  0.096  &  11.13  &  0  &  42.15  &  10.98  &   8.41  &   0.34 \\
 194.680  &  28.910  &  122.0  &  -2.0  &   5.350  &  0.090  &  12.42  &  1  &  42.08  &  10.55  &   8.90  &   0.41 \\
 193.922  &  27.251  &  103.5  &  -2.2  &   0.747  &  0.158  &  10.77  &  0  &  41.08  &  11.07  &   9.43  &   0.93 \\
 194.655  &  27.177  &  111.6  &  -1.2  & 39.096  &  0.323  &  12.65  &  1  &  42.86  &  10.38  &   6.47  &   0.06 \\
 194.638  &  27.364  &  101.7  &  -1.0  &   0.455  &  0.094  &  12.11  &  0  &  40.85  &  10.52  &  11.16  &   3.04 \\
 194.647  &  27.265  &  107.2  &  -1.8  & 24.348  &  0.101  &  13.43  &  1  &  42.62  &  10.04  &  \ldots  &  \ldots \\
 194.025  &  27.678  &    71.9  &  -2.0  & 67.728  &  0.105  &  12.93  &  1  &  42.72  &    9.89  &  \ldots  &  \ldots \\
 194.083  &  27.751  &  100.1  &  -5.0  &   0.576  &  0.102  &  11.57  &  0  &  40.94  &  10.72  &  10.00  &   0.93 \\
 194.269  &  27.773  &  109.9  &  -1.2  &   6.563  &  0.100  &  12.92  &  1  &  42.07  &  10.26  &  \ldots  &  \ldots \\
 193.854  &  27.798  &  107.1  &  -3.6  &   1.201  &  0.469  &  10.94  &  0  &  41.31  &  11.03  &    8.63  &   0.46 \\
 194.111  &  27.831  &    91.4  &  -5.0  &   0.815  &  0.087  &   11.51  &  0  &  41.01  &  10.67  &  10.72  &   2.21 \\
 194.320  &  27.618  &  105.6  &  -4.1  &   0.424  &  0.078  &  12.46  &  0  &  40.85  &  10.41  &  \ldots  &  \ldots \\
 194.717  &  27.785  &    82.4  &  -2.2  &   0.757  &  0.093  &  11.63  &  0  &  40.88  &  10.53  &   9.41  &   0.71 \\
 194.697  &  27.675  &  122.2  &  -2.1  &   0.470  &  0.367  &  10.93  &  0  &  41.02  &  11.15  &  \ldots  &  \ldots \\
 194.324  &  27.811  &  103.0  &  -2.3  &   0.309  &  0.137  &  12.74  &  0  &  40.69  &  10.28  &  13.21  &  10.00 \\
 194.742  &  27.595  &    87.9  &  -2.5  &   0.200  &  0.096  &  12.05  &  0  &  40.36  &  10.42  &  12.90  &  10.00 \\
 194.781  &  27.768  &    92.0  &  -4.4  &   0.789  &  0.134  &  11.81  &  0  &  41.00  &  10.55  &  11.57  &   5.41 \\
 194.784  &  27.784  &  103.1  &  -4.9  &   2.207  &  0.139  &  11.09  &  0  &  41.54  &  10.94  &  \ldots  &  \ldots \\
 194.647  &  27.596  &  111.8  &  -2.8  & 42.967  &  0.097  &  10.57  &  3  &  42.90  &  11.22  &    6.20  &   0.04 \\
 194.793  &  27.620  &    82.5  &  -4.1  &   0.771  &  0.098  &  12.15  &  0  &  40.89  &  10.32  &  10.17  &   1.10 \\
 194.946  &  27.710  &  122.9  &  -1.8  &   1.786  &  0.319  &  10.98  &  0  &  41.60  &  11.14  &  \ldots  &  \ldots \\
 195.301  &  27.604  &  110.2  &  -2.5  &   0.635  &  0.096  &  12.70  &  0  &  41.06  &  10.35  &  \ldots  &  \ldots \\
 195.474  &  27.624  &  114.9  &  -3.1  &   2.705  &  0.322  &    9.89  &  0  &  41.73  &  11.51  &    8.68  &   0.76 \\
 195.533  &  27.648  &  100.3  &  -1.9  & 63.709  &  0.098  &  11.74  &  1  &  42.98  &  10.65  &    5.84  &   0.03 \\
 194.806  &  27.775  &    99.6  &  -2.2  &   0.712  &  0.134  &  11.89  &  0  &  41.02  &  10.59  &  \ldots  &  \ldots \\
 194.453  &  28.180  &  105.3  &  -5.0  &   0.312  &  0.098  &  11.74  &  0  &  40.71  &  10.70  &  \ldots  &  \ldots \\
 194.375  &  28.188  &  100.8  &  -0.9  &   2.957  &  0.319  &  11.35  &  3  &  41.65  &  10.82  &  \ldots  &  \ldots \\
 194.162  &  28.081  &  118.5  &  -2.5  &   0.268  &  0.087  &  13.19  &  0  &  40.75  &  10.22  &  \ldots  &  \ldots \\
 194.558  &  28.183  &  104.5  &  -2.3  &   8.439  &  0.479  &  12.28  &  0  &  42.14  &  10.48  &    9.73  &   1.06 \\
 194.766  &  28.124  &  116.3  &  -4.6  &   4.750  &  0.326  &  10.29  &  0  &  41.98  &  11.36  &    8.26  &   0.30 \\
 194.590  &  28.149  &  114.6  &  -2.5  &   0.900  &  0.098  &  11.17  &  0  &  41.25  &  11.00  &  \ldots  &  \ldots \\
 194.652  &  28.114  &    98.7  &  -2.4  &   1.452  &  0.094  &  11.27  &  0  &  41.32  &  10.83  &    8.73  &   0.52 \\
 194.881  &  28.047  &  101.5  &  -3.4  &   0.857  &  0.103  &  11.80  &  0  &  41.12  &  10.64  &  \ldots  &  \ldots \\
 194.983  &  28.035  &  119.3  &  -2.6  &   1.813  &  0.148  &  11.16  &  0  &  41.59  &  11.04  &  10.20  &   1.44 \\
 194.861  &  27.999  &    97.1  &  -3.1  &   0.511  &  0.082  &  13.03  &  0  &  40.86  &  10.11  &    9.28  &   1.87 \\
 194.712  &  28.084  &    88.5  &  -2.3  &   0.270  &  0.095  &  12.48  &  0  &  40.50  &  10.25  &    9.76  &   0.84 \\
 194.833  &  28.084  &    68.0  &  -4.4  &   5.196  &  0.100  &  10.48  &  0  &  41.55  &  10.82  &    7.93  &   0.23 \\
 195.033  &  28.079  &  105.8  &  -4.8  &   2.033  &  0.097  &  11.59  &  0  &  41.53  &  10.76  &    9.87  &   0.90 \\
 195.027  &  28.004  &  105.8  &  -4.7  &   0.921  &  0.091  &  12.14  &  0  &  41.19  &  10.54  &  10.64  &   2.38 \\
 195.054  &  28.076  &  109.3  &  -3.3  &   0.570  &  0.100  &  12.26  &  0  &  41.01  &  10.52  &    9.42  &   0.81 \\
 195.071  &  28.064  &    89.3  &  -2.5  &   0.716  &  0.084  &  12.40  &  0  &  40.93  &  10.29  &  10.60  &   1.72 \\
 195.092  &  28.047  &  119.6  &  -2.4  &   0.729  &  0.098  &  11.55  &  0  &  41.19  &  10.88  &    9.74  &   1.05 \\
 195.186  &  28.101  &    96.1  &  -2.3  &   0.408  &  0.094  &  12.55  &  0  &  40.75  &  10.30  &  \ldots  &  \ldots \\
 195.203  &  28.091  &  100.9  &  -4.2  &   2.282  &  0.094  &  10.90  &  3  &  41.54  &  11.00  &   9.30  &   0.63 \\
 195.038  &  28.170  &    98.6  &  -4.7  &   0.781  &  0.138  &  11.77  &  0  &  41.05  &  10.63  &  10.16  &   1.34 \\
 195.227  &  28.008  &    72.3  &  -4.6  &   2.903  &  0.333  &  10.41  &  0  &  41.35  &  10.90  &    9.29  &   1.03 \\
 195.061  &  28.041  &    83.2  &  -4.9  &   1.953  &  0.096  &  11.78  &  0  &  41.30  &  10.48  &  10.29  &   1.46 \\
 195.170  &  27.997  &  103.5  &  -4.6  &   0.452  &  0.101  &  11.66  &  0  &  40.86  &  10.71  &  10.20  &   1.30 \\
 195.363  &  27.999  &  111.5  &  -4.1  &   0.511  &  0.076  &  12.63  &  0  &  40.98  &  10.39  &  \ldots  &  \ldots \\
 195.446  &  28.095  &    84.3  &  -2.8  &   1.137  &  0.101  &  11.20  &  0  &  41.08  &  10.72  &  \ldots  &  \ldots \\
 195.490  &  28.006  &  113.1  &  -3.1  &   5.693  &  0.345  &  10.36  &  0  &  42.04  &  11.31  &    8.03  &   0.21 \\
 195.754  &  28.032  &    78.8  &  -1.9  &   2.114  &  0.080  &  10.31  &  0  &  41.29  &  11.02  &    8.28  &   0.25 \\
 195.736  &  28.070  &  113.0  &  -2.3  &   0.244  &  0.098  &  12.13  &  0  &  40.67  &  10.60  &  \ldots  &  \ldots \\
 195.818  &  28.030  &    87.6  &  -0.5  &   9.402  &  0.321  &  11.15  &  3  &  42.03  &  10.77  &    7.24  &   0.12 \\
 195.937  &  28.084  &    81.2  &  -2.0  &   6.821  &  0.100  &  11.40  &  3  &  41.83  &  10.61  &    8.18  &   0.15 \\
 194.402  &  27.031  &  106.8  &  -0.2  &   0.774  &  0.132  &  11.69  &  0  &  41.12  &  10.73  &  \ldots  &  \ldots \\
 194.300  &  27.103  &  105.8  &  -2.0  &   0.325  &  0.103  &  13.90  &  3  &  40.74  &    9.84  &  \ldots  &  \ldots \\
 194.181  &  27.179  &  111.1  &  -3.0  &   4.747  &  0.352  &  10.04  &  0  &  41.94  &  11.42  &    9.02  &   0.81 \\
 194.289  &  27.466  &  108.7  &  -2.1  &   1.379  &  0.102  &  11.11  &  0  &  41.39  &  10.98  &    9.09  &   0.65 \\
 194.177  &  27.548  &  110.3  &  -2.0  &   0.243  &  0.144  &  13.89  &  0  &  40.64  &    9.88  &  \ldots  &  \ldots \\
 194.399  &  27.493  &  106.9  &  -4.7  &   1.298  &  0.098  &  10.99  &  0  &  41.35  &  11.01  &    9.21  &   0.54 \\
 194.401  &  27.485  &  105.3  &  -4.7  &   0.576  &  0.094  &  12.14  &  0  &  40.98  &  10.54  &    9.56  &   0.71 \\
 194.295  &  27.405  &    90.2  &  -1.0  &   5.039  &  0.103  &  12.10  &  3  &  41.79  &  10.42  &    8.71  &   0.32 \\
 194.070  &  27.446  &    92.7  &  -4.0  &   0.236  &  0.095  &  12.12  &  0  &  40.48  &  10.43  &  10.00  &   0.93 \\
 194.358  &  27.546  &    79.3  &  -2.4  &   0.746  &  0.081  &  11.84  &  0  &  40.85  &  10.41  &  \ldots  &  \ldots \\
 194.355  &  27.405  &    70.2  &  -2.5  & 14.344  &  0.099  &  12.78  &  2  &  42.02  &    9.93  &  \ldots  &  \ldots \\
 194.484  &  27.581  &    71.7  &  -0.4  &   1.851  &  0.091  &  14.42  &  1  &  41.15  &    9.29  &  \ldots  &  \ldots \\
 194.143  &  27.539  &  103.3  &  -4.4  &   0.743  &  0.120  &  11.60  &  0  &  41.07  &  10.74  &  10.52  &   1.61 \\
 194.525  &  27.419  &    83.3  &  -3.0  &   0.513  &  0.094  &  13.45  &  0  &  40.73  &    9.81  &  \ldots  &  \ldots \\
 194.658  &  27.464  &    91.5  &  -2.5  &   5.507  &  0.104  &  12.88  &  1  &  41.84  &  10.12  &  \ldots  &  \ldots \\
 194.586  &  27.429  &  110.4  &  -3.0  &   1.714  &  0.134  &  14.36  &  1  &  41.49  &    9.69  &  \ldots  &  \ldots \\
 194.634  &  27.456  &  102.6  &  -2.5  &   0.497  &  0.104  &  12.44  &  0  &  40.89  &  10.39  &  11.08  &   4.23 \\
 195.080  &  27.554  &    85.6  &  -3.1  &   0.583  &  0.103  &  11.84  &  0  &  40.80  &  10.48  &  \ldots  &  \ldots \\
 194.807  &  27.403  &    82.3  &  -1.9  &   1.650  &  0.084  &  11.44  &  0  &  41.22  &  10.60  &    9.85  &   0.95 \\
 193.989  &  27.905  &    95.2  &  -5.0  &   0.553  &  0.091  &  13.30  &  0  &  40.87  &    9.99  &  \ldots  &  \ldots \\
 194.179  &  28.020  &    91.4  &  -1.9  &   1.764  &  0.335  &  10.92  &  0  &  41.34  &  10.90  &    9.38  &   0.68 \\
 194.341  &  27.880  &  108.1  &  -4.4  &   0.455  &  0.103  &  12.79  &  0  &  40.90  &  10.30  &  \ldots  &  \ldots \\
 194.450  &  27.883  &    85.6  &  -4.7  &   1.665  &  0.103  &  11.40  &  0  &  41.26  &  10.65  &  10.64  &   2.78 \\
 194.447  &  27.833  &    88.8  &  -2.3  &   0.610  &  0.099  &  12.02  &  0  &  40.86  &  10.43  &  \ldots  &  \ldots \\
 194.515  &  27.815  &  103.7  &  -2.3  &   0.575  &  0.102  &  11.98  &  0  &  40.97  &  10.59  &  11.84  &   7.52 \\
 194.591  &  27.968  &    87.4  &  -2.3  &   5.115  &  0.101  &  11.03  &  3  &  41.77  &  10.82  &    8.84  &   0.55 \\
 194.703  &  27.810  &    85.0  &  -2.2  &   1.586  &  0.095  &  11.12  &  0  &  41.23  &  10.76  &  10.00  &   1.36 \\
 194.814  &  27.971  &    69.2  &  -4.9  &   2.589  &  0.094  &  11.00  &  0  &  41.27  &  10.63  &    8.97  &   0.46 \\
 194.783  &  27.855  &    96.2  &  -2.2  &   2.779  &  0.103  &  11.50  &  3  &  41.58  &  10.71  &    9.48  &   0.80 \\
 194.769  &  27.911  &    93.4  &  -2.0  &   0.255  &  0.098  &  12.66  &  0  &  40.52  &  10.23  &  10.78  &   2.15 \\
 194.834  &  27.886  &    93.7  &  -3.4  &   0.583  &  0.099  &  12.46  &  0  &  40.88  &  10.31  &  10.35  &   1.53 \\
 194.750  &  27.968  &  121.5  &  -4.4  &   0.218  &  0.100  &  12.89  &  0  &  40.68  &  10.36  &  \ldots  &  \ldots \\
 194.775  &  27.997  &  111.8  &  -2.1  &   1.739  &  0.098  &  11.38  &  0  &  41.51  &  10.89  &   9.52  &   0.97 \\
 194.733  &  27.833  &  110.3  &  -1.6  & 16.290  &  0.330  &  11.01  &  3  &  42.47  &  11.03  &  \ldots  &  \ldots \\
 194.855  &  27.968  &  112.1  &  -3.3  &   0.217  &  0.094  &  12.61  &  0  &  40.61  &  10.40  &  12.10  &   7.51 \\
 194.908  &  27.907  &  117.3  &  -2.2  &   2.139  &  0.146  &  11.56  &  0  &  41.64  &  10.86  &  10.15  &   1.18 \\
 194.887  &  27.984  &    83.9  &  -2.1  &   1.395  &  0.098  &  11.25  &  0  &  41.17  &  10.70  &   9.48  &   0.76 \\
 194.935  &  27.913  &    97.0  &  -4.7  &   1.129  &  0.095  &  11.34  &  0  &  41.20  &  10.79  &   9.35  &   0.65 \\
 194.942  &  27.857  &  118.4  &  -3.3  &   0.152  &  0.096  &  12.04  &  0  &  40.50  &  10.68  &  \ldots  &  \ldots \\
 194.899  &  27.959  &  104.4  &  -3.6  & 25.884  &  0.492  &    8.86  &  0  &  42.62  &  11.84  &  \ldots  &  \ldots \\
 194.872  &  27.850  &    99.8  &  -4.2  &   3.066  &  0.152  &  11.50  &  0  &  41.66  &  10.75  &  \ldots  &  \ldots \\
 194.878  &  27.884  &    68.8  &  -1.9  &   3.057  &  0.143  &  11.10  &  0  &  41.33  &  10.58  &   9.28  &   0.81 \\
 194.986  &  27.930  &  112.7  &  -2.3  &   0.228  &  0.080  &  12.62  &  0  &  40.64  &  10.40  &  \ldots  &  \ldots \\
 195.166  &  27.924  &  109.5  &  -4.9  &   0.505  &  0.104  &  11.24  &  0  &  40.96  &  10.93  &   8.87  &   0.64 \\
 195.026  &  27.776  &    90.2  &  -3.3  &   0.464  &  0.104  &  12.04  &  0  &  40.75  &  10.44  & 10.22  &   1.35 \\
 195.023  &  27.808  &    95.8  &  -2.5  &   0.706  &  0.098  &  11.86  &  0  &  40.99  &  10.56  &  \ldots  &  \ldots \\
 195.135  &  27.766  &    97.3  &  -1.5  &   0.538  &  0.099  &  12.78  &  0  &  40.88  &  10.21  &   9.26  &   0.55 \\
 195.034  &  27.977  &    94.3  &  -4.3  &   6.201  &  0.346  &    8.41  &  0  &  41.92  &  11.93  &   7.20  &   0.62 \\
 195.178  &  27.971  &    92.8  &  -1.9  &   4.849  &  0.099  &  10.76  &  3  &  41.79  &  10.98  &   8.22  &   0.33 \\
 195.323  &  27.809  &  106.7  &  -1.9  &   1.474  &  0.133  &  10.85  &  0  &  41.40  &  11.07  &  \ldots  &  \ldots \\
 195.383  &  27.847  &    79.4  &  -3.1  &   0.638  &  0.345  &  10.72  &  0  &  40.78  &  10.86  &  \ldots  &  \ldots \\
 195.633  &  27.936  &    96.4  &  -2.5  &   0.878  &  0.091  &  13.39  &  0  &  41.09  &    9.96  &   9.45  &   0.53 \\
 195.459  &  27.894  &  110.0  &  -4.2  &   1.446  &  0.096  &  11.37  &  0  &  41.42  &  10.88  & 10.35  &   1.54 \\
 195.501  &  27.783  &  102.8  &  -2.5  &   1.014  &  0.097  &  12.34  &  0  &  41.20  &  10.44  & 10.21  &   1.39 \\
 195.720  &  27.867  &  119.8  &  -0.6  &   0.816  &  0.098  &  11.28  &  0  &  41.24  &  10.99  &  11.75  &   5.51 \\
 194.623  &  28.301  &    85.3  &  -2.5  &   2.180  &  0.115  &  13.34  &  0  &  41.37  &    9.88  &  \ldots  &  \ldots \\
 195.075  &  28.202  &  123.9  &  -2.1  &   3.111  &  0.492  &  10.14  &  0  &  41.85  &  11.48  &    7.81  &   0.23 \\
 195.128  &  28.346  &    87.2  &  -3.0  &   2.760  &  0.346  &  10.76  &  0  &  41.50  &  10.92  &   9.34  &   0.95 \\
 195.184  &  28.337  &  115.6  &  -2.3  &   0.709  &  0.145  &  11.74  &  0  &  41.15  &  10.78  &  \ldots  &  \ldots \\
 194.758  &  28.225  &  116.9  &  -1.4  &   6.640  &  0.101  &  11.04  &  3  &  42.13  &  11.07  &    8.45  &   0.30 \\
 194.683  &  28.283  &  110.6  &  -2.5  &   0.052  &  0.098  &  13.76  &  0  &  39.98  &    9.93  &  \ldots  &  \ldots \\
 195.543  &  28.192  &   83.2  &  -3.4  &   1.312  &  0.142  &  12.83  &  0  &  41.13  &  10.06  &  11.42  &   2.98 \\
 195.590  &  28.231  &   80.9  &  -1.9  &   0.949  &  0.144  &  11.57  &  0  &  40.97  &  10.54  &    9.40  &   0.89 \\
 195.590  &  28.256  &   99.0  &  -2.3  &   3.463  &  0.139  &  11.88  &  3  &  41.71  &  10.59  &    9.02  &   0.49 \\
 196.042  &  28.248  &   86.4  &  -3.1  &   2.378  &  0.099  &  11.47  &  0  &  41.42  &  10.63  &  10.81  &   1.64
\enddata 
\tablecomments{Col. (1) \& (2): J2000 epoch Right Ascension and Declination as listed in \citet{Mahajan:10}. 
Col. (3) Adopted distance, converted from the redshift measurements from \citet{Mahajan:10}, using $H_0 = 70$ km s$^{-1}$ 
Mpc$^{-1}$.
Col. (4): Morphological T parameter obtained from the Hyperleda database \citep{Paturel:03}.
Col. (5) \& (6): {\em Spitzer} MIPS 24$\mu$m flux densities and associated uncertainties from \citet{Mahajan:10}. The values in Col. 
(6) were provided to us by private communication from Mahajan.
Col. (7): K$_s$ magnitude from 2MASS.
Col. (8): AGN classification from \citet{Mahajan:10}:  unclassified source are flagged as 0, 1 for star forming galaxies, $\ge$ 2 
indicates an AGN.
Col. (9): Log of the 24$\mu$m luminosity.
Col. (10): Log of the Ks-band luminosity in solar units.
Col. (11) \& (12): WISE 22$\mu$m magnitude and associated error. 
}
\end{deluxetable*}
}

\bibliography{biblio}

\end{document}